\documentclass[twocolumn,epsfig,graphics,floatfix,nofootinbib]{revtex4-1}

\usepackage[table]{xcolor}
\usepackage{amsmath,amsfonts,amssymb,graphics,graphicx,epsfig,color,times,bbm}
\usepackage{amsthm}
\usepackage{psfrag}
\usepackage[hidelinks]{hyperref}
\hypersetup{colorlinks=false}
\usepackage{wasysym}
\usepackage{bbm}

\usepackage{hyperref}
\usepackage{lipsum}
\usepackage{booktabs}

\begin{document}

\title{Interleaving: Modular architectures for fault-tolerant photonic quantum computing}

\newcommand*\leadauthor{\thanks{These authors contributed equally:  \\ \mbox{dlitinski@psiquantum.com} \\  \mbox{naomi@psiquantum.com} \\ \mbox{fernando@psiquantum.com}}}

\author{H\'ector Bomb\'in}
\author{Isaac H. Kim}
\author{Daniel Litinski} \leadauthor
\author{Naomi Nickerson} \leadauthor
\author{Mihir Pant}
\author{Fernando Pastawski} \leadauthor
\author{Sam Roberts}
\author{Terry Rudolph}

\affiliation{PsiQuantum, Palo Alto}
    
\cleardoublepage

\begin{abstract}

Useful fault-tolerant quantum computers require very large numbers of physical qubits. 
Quantum computers are often designed as arrays of static qubits executing gates and measurements.
Photonic qubits require a different approach. 
In photonic fusion-based quantum computing (FBQC), the main hardware components are resource-state generators (RSGs) and fusion devices connected via waveguides and switches. 
RSGs produce small entangled states of a few photonic qubits, whereas fusion devices perform entangling measurements between different resource states, thereby executing computations. 
In addition to these components, low-loss photonic delays such as optical fiber can be used as fixed-time quantum memories simultaneously storing thousands of photonic qubits. 
Here, we present a modular architecture for FBQC in which these components are combined to form \textit{interleaving modules} consisting of one RSG with its associated fusion devices and a few fiber delays. 
Exploiting the multiplicative power of delay components, each interleaving module can add thousands of physical qubits to the computational Hilbert space.
Networks of interleaving modules are universal fault-tolerant quantum computers, which we demonstrate using surface codes and lattice surgery as a guiding example.
Our numerical analysis shows that in a network of modules containing 1-km-long fiber delays, a single RSG can generate four logical surface-code qubits with a code distance of 35 while tolerating photon loss rates above 2\% in addition to the fiber-delay loss.
We illustrate how the combination of interleaving with further uses of non-local fiber connections can reduce the cost of various logical operations and facilitate the implementation of unconventional geometries such as periodic boundaries or stellated surface codes.
Interleaving applies beyond purely optical architectures, and can also turn many small disconnected matter-qubit devices with transduction to photons into a large-scale quantum computer.

\end{abstract}

\maketitle

\textbf{Photons are different.} Solid-state architectures for large-scale quantum computing are usually described as arrays of static qubits that store quantum information and are capable of performing long sequences of gates and measurements. 
When constructing devices that use photons as qubits, such a description is not suitable.
Single photons are short-lived, easily lost and destroyed after measurement, making them difficult to use as static qubits. 
An approach that is better suited for fault-tolerant photonic quantum computing is \textit{fusion-based quantum computing} (FBQC)~\cite{FBQCpaper}.
In this work, we introduce a modular architecture for fault-tolerant FBQC. 

In photonic FBQC, a large-scale quantum computer is not an array of static qubits, but rather a network of \textit{resource-state generators} (RSGs) and \textit{fusion devices}. 
Each RSG is a device that periodically produces entangled few-photon resource states. 
Waveguides transport these photons into fusion devices that perform \textit{fusions}~\cite{Browne2005, GimenoSegovia2015}, which are entangling few-photon measurements between different resource states. 
Even though the measured photons are destroyed, their quantum information is preserved, as it is teleported from resource state to resource state through these measurements. 
In this approach, quantum information is not stored in a static array of qubits, but periodically teleported from older resource states to freshly generated ones.

How does one then compare the computational power of static matter-based qubits to that of photonic qubits?
Matter-based architectures are typically characterized by the number of physical qubits and the speed of physical gates and measurements.
In FBQC, the physical qubits are short-lived photonic degrees of freedom.
In the simplest case, these consist of a single photon (e.g., dual-rail qubits), but can also consist of multiple photons (e.g., GKP qubits~\cite{Gottesman2001}).
In the dual-rail case, the qubit count is proportional to the number of photons simultaneously present in the device. 
The computational speed is determined by the rate of physical entangling measurements, i.e., fusions of resource states. 
More pragmatically, in fault-tolerant quantum computation~\cite{Preskill1998,Terhal2015,Campbell2016}, we consider a static qubit-based device and a dynamic RSG-based device equally powerful, if they can execute the same quantum computation in the same amount of time.

\textbf{Fiber as memory.} One remarkable advantage of photonic architectures is the availability of high-capacity quantum memory. 
Optical fiber features exceptionally low transmission loss rates of less than 0.2~dB/km at telecom wavelengths (1550 nm)~\cite{Li2020}. 
In other words, a photon entering a 1-km-long low-loss fiber has a $> \! 95\%$ probability of exiting the fiber a few microseconds later, which, as we show, is a loss rate compatible with fault-tolerant FBQC. 
If one photon enters this fiber every nanosecond, the fiber then acts as a temporary memory for up to 5,000 photons. 
However, photons can only be stored in fiber for a fixed amount of time and cannot be accessed until they leave the fiber.
We introduce an architecture in which, despite these limitations, each RSG together with its associated fusion devices and such a memory can provide the computational power of thousands of physical qubits for the purpose of fault-tolerant quantum computing.
For example, we demonstrate that in a network of RSGs equipped with 5,000-photon delays, each RSG adds approximately four logical surface-code qubits with a code distance of 35. Importantly, this network supports a universal set of logical gates. In an array of static qubits, an equivalent implementation would require 5,000 physical data qubits per RSG.

Previous studies considered the use of fiber as photonic memory~\cite{Roncaglia2011, Rohde2015, Pichler2017, Asavanant2019, Wan2020} and the use of memory to increase the number of qubits in fault-tolerant quantum computation~\cite{Wan2020, Duckering2020}. 
Here, we study the interplay of photonic FBQC, fiber as memory and topological fault-tolerance protocols. 
Our main goal is to demonstrate the following:

\begin{enumerate}
\item \textbf{A single RSG is much more powerful than a single static qubit.} 
By temporarily storing photonic resource states in a low-loss medium, such as optical fiber, an RSG can produce multiple~--~~potentially thousands of~--~simultaneously existing resource states.
This enables each RSG to emulate thousands of static physical qubits for the purpose of fault-tolerant quantum computing.
\item \textbf{Architectures for photonic FBQC are highly modular and scalable.} 
Large-scale fault-tolerant quantum computers can be constructed as networks of identical modules.
Each module consists of a single RSG as its main component, a few fusion devices and macroscopic fiber delays that are used as memory and as connections between modules.
When scaling up such a quantum computer, the main challenge is therefore the construction of many identical RSGs, rather than a large array of static qubits.
While RSGs are primarily envisioned as linear-optical devices, the architecture is agnostic to the inner workings of an RSG.
This provides an alternative approach to scale up non-photonic matter-based devices, such as solid-state qubits, by using them as autonomously operating RSGs embedded in a large-scale photonic architecture, provided that transduction to suitable photons is available.
\item \textbf{Macroscopic optical connections between modular components can reduce the cost of logical operations.} 
The photons produced by RSGs may travel large distances and are not affected by the same locality constraints as conventional architectures. Non-local connections between RSGs provide a new set of tools to implement logical operations more efficiently, which we demonstrate using the examples of periodic boundary conditions, qubit routing and the implementation of unconventional geometries such as stellated surface codes.

\end{enumerate}

The remainder of this paper is structured as follows. In Sec.~\ref{sec:principles}, we review FBQC using the example of 6-ring resource states~\cite{FBQCpaper}.
We show how to construct interleaving modules to effectively increase the number of available physical qubits. 
In Sec.~\ref{sec:logic}, we review how to use 6-ring resource states for universal fault-tolerant quantum computation using the example of surface codes~\cite{Kitaev2003, Bravyi1998, Fowler2012} and lattice surgery~\cite{Horsman2012, Fowler2018, Litinski2019}, and show explicit constructions of networks of interleaving modules that can be used for universal quantum computing.
In Sec.~\ref{sec:architecture}, we present numerical results for the loss tolerance of this architecture.
Using a few examples, we demonstrate that fiber connections between modular hardware components significantly facilitate the implementation of exotic (and potentially more efficient) logical protocols which are difficult to implement with 2D arrays of static physical qubits.
We also discuss how interleaving can not only be used in purely photonic architectures, but also has the potential to scale up disconnected few-qubit matter-based devices by using them as RSGs, provided they are capable of transduction to suitable photons.

\begin{figure*}[t]
\centering
\includegraphics[width=\linewidth]{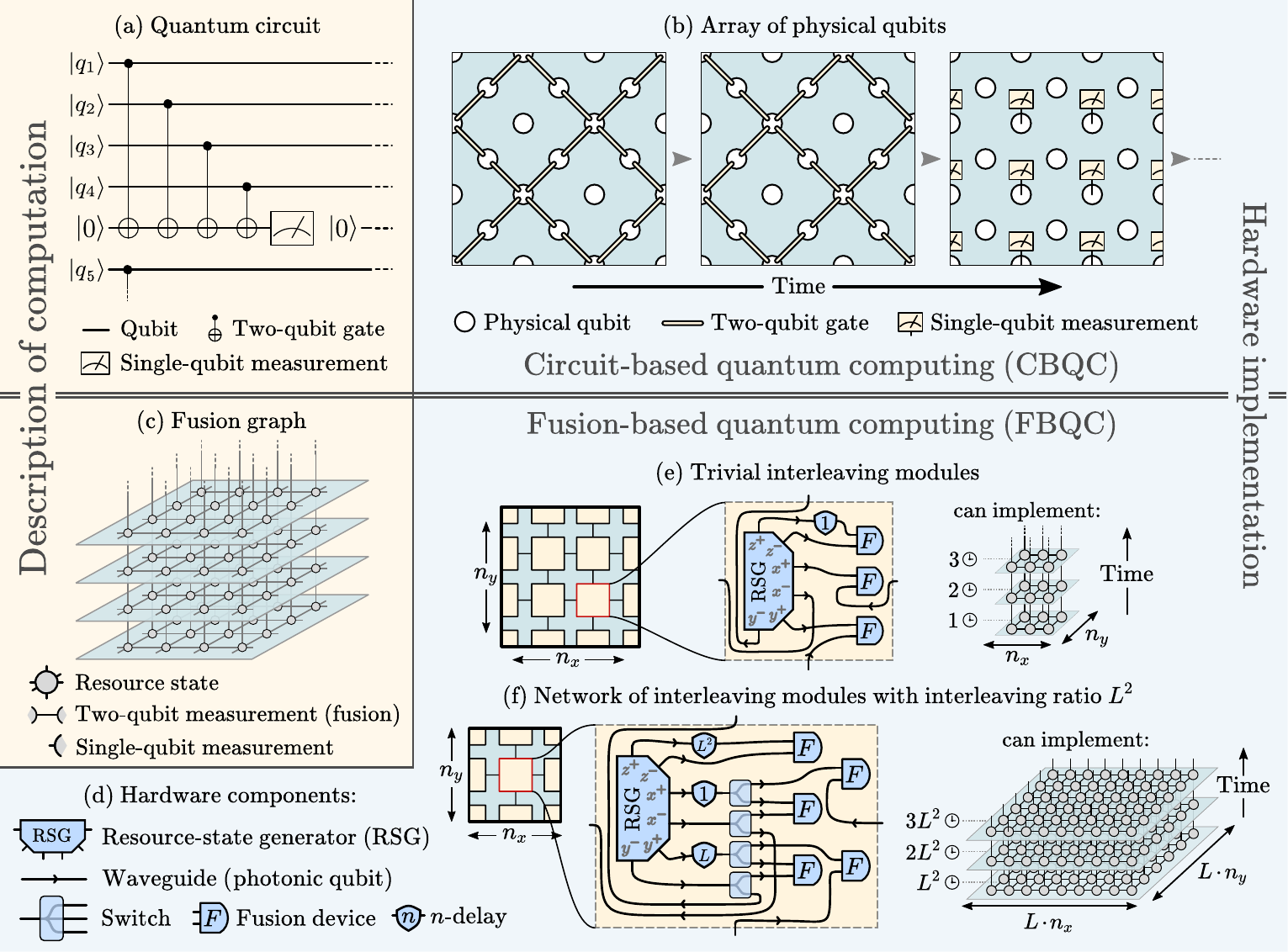}
\caption{
Overview of FBQC and interleaving. 
(a) In circuit-based quantum computing (CBQC), computations are described by quantum circuits, typically containing two-qubit gates and measurements for error correction. 
(b) These are implemented by consecutively executing layers of gates and measurements in arrays of physical qubits. For error correction with topological codes such as surface codes, these qubits are typically arranged on 2D grids, and all physical gates involve only neighboring qubits. 
(c) A fusion-based quantum computation is described by a fusion graph, in which each vertex is a few-qubit resource state, each edge a two-qubit fusion instruction, and each half-edge a single-qubit measurement. In fault-tolerant quantum computation with topological codes, fusion graphs are typically 2+1-dimensional, where 2D slices roughly correspond to layers of gates and measurements executed by a 2D array of qubits in CBQC. 
(d) The hardware components in FBQC are resource-state generators (RSGs) producing resource states at periodic time intervals called \textit{RSG cycles} (\clock), fusion devices performing single- and two-qubit measurements and $n$-delays that store photons for $n${\clock}. 
These components are connected via waveguides and switches that transport and reroute photons. 
(e) 2D arrays of $n_x \times n_y$ modules containing an RSG and a 1{\clock}-delay can implement fusion graphs with 2D slices of size $n_x \times n_y$. 
(f) Augmenting these modules with an $L${\clock}-delay, an $L^2${\clock}-delay and a few switches turns them into \textit{interleaving modules} capable of producing fusion graphs with $L^2$ times larger 2D slices, but $L^2$ times more slowly, effectively slowing down the generation of slices in exchange for more physical qubits. 
Values of $L^2>1000$ are achievable with realistic physical components.
}

\label{fig:overview}
\end{figure*}

\definecolor{col2}{HTML}{d2e6ea}
\def\arraystretch{1.5}
\begin{table*}
\rowcolors{1}{col2}{gray!10}
\scalebox{1}{
\begin{tabular}{p{\linewidth}}

\quad
\begin{tabular}{p{0.96\linewidth}}
\textbf{Fusion-based quantum computing (FBQC) \, $|$ \,} Approach to quantum computing in which all operations are performed by the generation of few-qubit resource states and few-qubit entangling measurements. \end{tabular} \\

\quad
\begin{tabular}{p{0.96\linewidth}}
\textbf{Photonic qubit \, $|$ \,} A qubit associated with the degrees of freedom of one or several photons. 
Examples include polarization, time-bin or dual-rail encoding of a single photon, and GKP bosonic encoding of multiple photons. 
Any of these photonic qubits can also be encoded in a stabilizer code such as a Shor code to define an encoded photonic qubit. \end{tabular} \\

\quad
\begin{tabular}{p{0.96\linewidth}}
\textbf{Resource state \, $|$ \,} Entangled few-qubit state. \end{tabular} \\

\quad
\begin{tabular}{p{0.96\linewidth}}
\textbf{Fusion \, $|$ \,} Entangling multi-qubit measurement. Here we only consider two-qubit measurements. \end{tabular} \\

\quad
\begin{tabular}{p{0.96\linewidth}}
\textbf{Resource-state generator (RSG) \, $|$ \,} Elementary hardware component. Generates one resource state in every RSG cycle. \end{tabular} \\

\quad
\begin{tabular}{p{0.96\linewidth}}
\textbf{RSG cycle $t_{\mathrm{RSG}}$ \, $|$ \,} Clock cycle in FBQC in which each RSG produces one resource state. Denoted by \clock~in the figures. \end{tabular} \\

\quad
\begin{tabular}{p{0.96\linewidth}}
\textbf{Fusion device \, $|$ \,} Elementary hardware component. Performs entangling few-qubit measurements on incoming photonic qubits. May be \textit{reconfigurable}, i.e., capable of performing different types of measurements on demand, including single-qubit measurements. \end{tabular} \\

\quad
\begin{tabular}{p{0.96\linewidth}}
\textbf{$n$-delay \, $|$ \,} Fixed-time quantum memory that stores each photonic qubit for exactly $n$ RSG cycles. \end{tabular} \\

\quad
\begin{tabular}{p{0.96\linewidth}}
\textbf{Fusion graph \, $|$ \,} Graph that describes which operations need to be performed in a fusion-based quantum computation. Each vertex is a resource state, each edge is a two-qubit fusion, and each half-edge connected to a single vertex is a single-qubit measurement. \end{tabular} \\

\quad
\begin{tabular}{p{0.96\linewidth}}
\textbf{Fusion-graph slice \, $|$ \,} For $D+1$-dimensional fusion graphs: a $D$-dimensional layer of the fusion graph. \end{tabular} \\

\quad
\begin{tabular}{p{0.96\linewidth}}
\textbf{Interleaving \, $|$ \,} The use of long $n$-delays to increase the size of fusion-graph slices that can be produced by a collection of RSGs. \end{tabular} \\

\quad
\begin{tabular}{p{0.96\linewidth}}
\textbf{Interleaving coordinates \, $|$ \,} Coordinates $(g,t)$ that are uniquely assigned to resource states in a fusion graph to be implemented by a network of RSGs. Here, $g$ labels the RSG that will produce the resource state, and $t$ the RSG cycle in which it will be produced. \end{tabular} \\

\quad
\begin{tabular}{p{0.96\linewidth}}
\textbf{Instantaneous / delayed / local / networked fusion \, $|$ \,} A fusion between two qubits from resource states with interleaving coordinates $(g_1,t_1)$ and $(g_2,t_2)$ is called instantaneous if $t_1 = t_2$, delayed if $t_1 \neq t_2$, local if $g_1 = g_2$ and networked if $g_1 \neq g_2$. \end{tabular} \\

\quad
\begin{tabular}{p{0.96\linewidth}}
\textbf{Interleaving module \, $|$ \,} Modular hardware component consisting of one RSG, different $n$-delays and fusion devices. Each photon produced by the RSG may enter a switch that, depending on the switch setting, routes the photon to one of several (reconfigurable) fusion devices, potentially with an additional delay. Quantum computation is performed by choosing switch settings in every RSG cycle. \end{tabular} \\

\quad
\begin{tabular}{p{0.96\linewidth}}
\textbf{Network of interleaving modules \, $|$ \,} Quantum computer consisting of $n_{\mathrm{RSG}}$ interleaving modules. Can execute quantum computations described by fusion graphs with interleaving coordinates $(g,t)$ where $g \leq n_{\mathrm{RSG}}$. Additional constraints on the interleaving coordinates are set by the available hardware components. Specifically, the connections between modules determine which networked fusions are allowed, while the available delays determine which delayed fusions are allowed. \end{tabular} \\

\quad
\begin{tabular}{p{0.96\linewidth}}
\textbf{Interleaving ratio \, $|$ \,} Ratio of $n_{\mathrm{slice}}$ to $n_{\mathrm{RSG}}$ quantifying the multiplicative power of interleaving. Here, $n_{\mathrm{slice}}$ is the size of fusion-graph slices (measured in number of resource states) that can be generated by a network of interleaving modules. For the modules discussed in Secs.~\ref{sec:principles} and \ref{sec:logic}, this ratio is $L^2$, where $L$ is a parameter called the \textit{rastering length}. \end{tabular} \\

\end{tabular}}
\caption{Glossary of terms in FBQC and interleaving.}
\label{tab:definitions}
\end{table*}

\begin{figure*}[t]
\centering
\includegraphics[width=\linewidth]{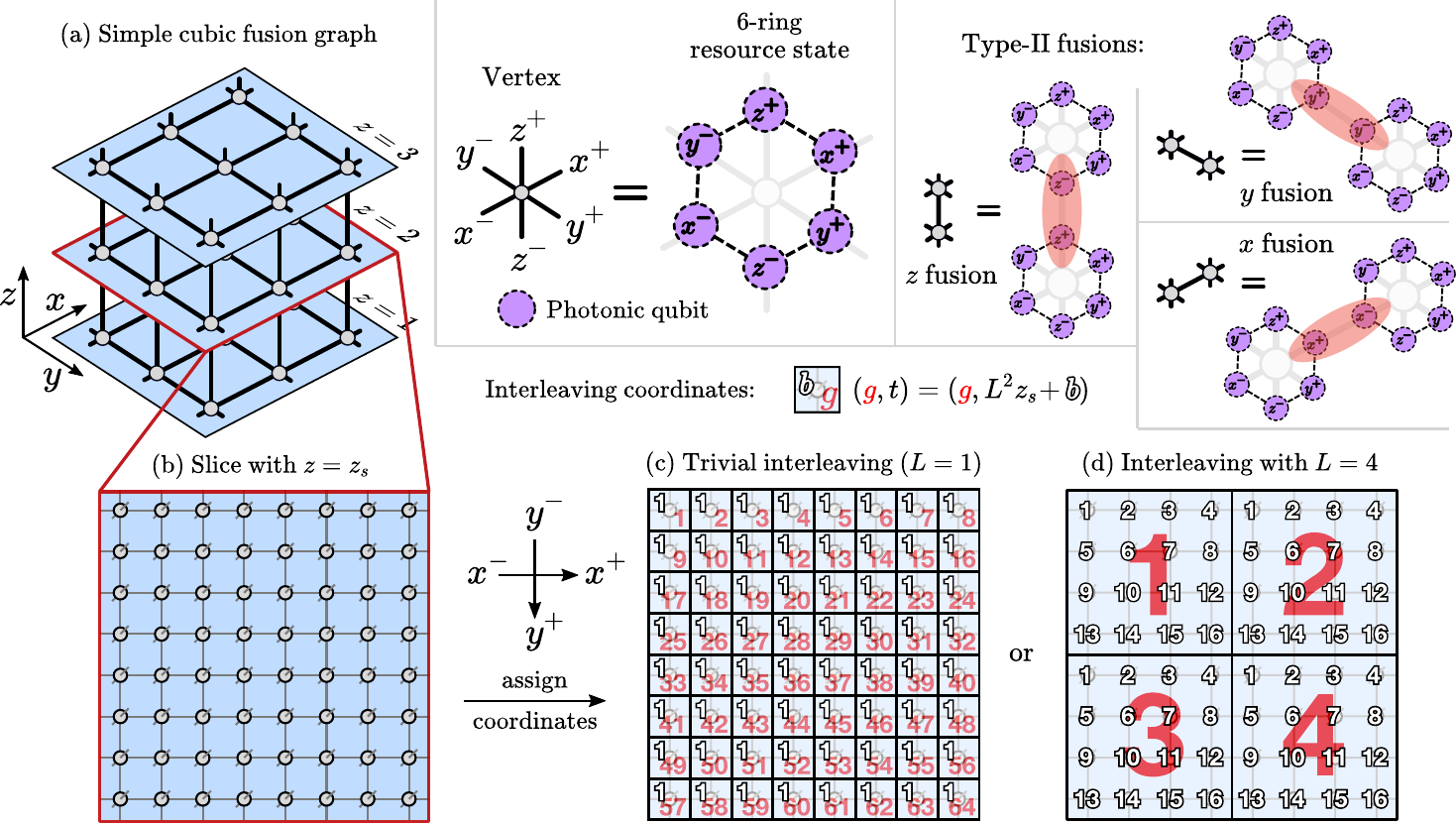}
\caption{
Simple cubic fusion graphs and assignment of interleaving coordinates. 
(a) The fusion graph is a simple cubic lattice in which each vertex corresponds to a 6-ring resource state. 
This resource state consists of six photonic qubits which are labeled $x^\pm$, $y^\pm$ and $z^\pm$, according to the six directions of the cubic fusion graph. 
Each edge of the graph corresponds to a fusion. 
(b) 6-ring fusion graphs can be partitioned into 2D slices perpendicular to the $z$ direction. 
In this example, each slice consists of 64 resource states. 
(c) Interleaving coordinates $(g,t)$ are assigned to each resource state in the fusion graph to determine the RSG $g$ that will produce this resource state and the RSG cycle $t$ during which it will be produced. 
In the trivial case, 64 RSGs are used to produce one 2D slice in every RSG cycle, which is compatible with the hardware modules shown in Fig.~\ref{fig:l1modules}a. 
(d) Alternatively, when interleaving with rastering length $L=4$, the fusions instructions of each 2D slice can be produced in 16 RSG cycles using only four RSGs. The hardware modules capable of these operations are shown in Fig.~\ref{fig:bulkmodule}.}
\label{fig:fusiongraph}
\end{figure*}

\section{Principles of FBQC and Interleaving}

\label{sec:principles}

\begin{figure*}
\centering
\includegraphics[width=0.9\linewidth]{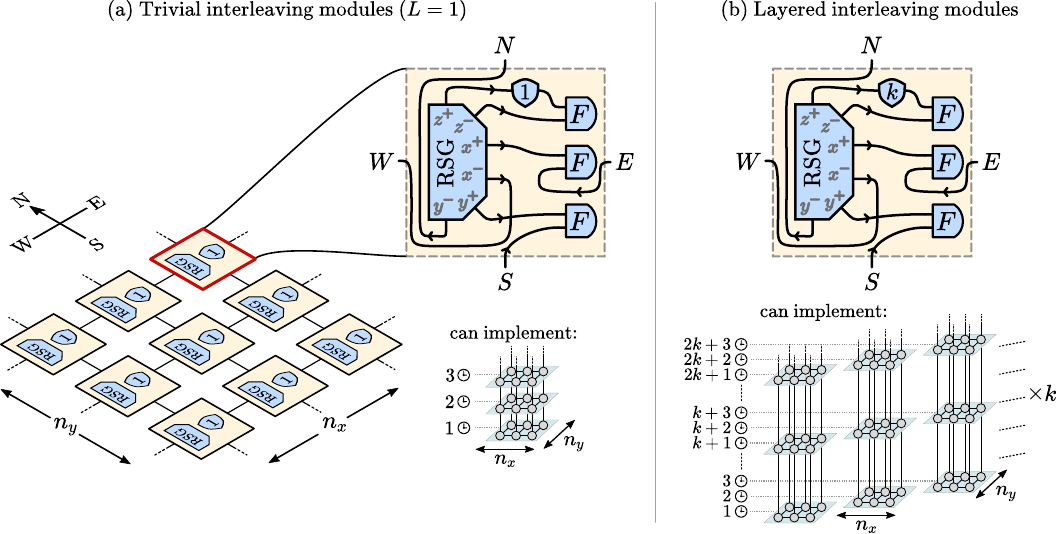}
\caption{(a) Network of modules capable of producing one 2D slice of size $n_x \times n_y$ of a cubic 6-ring fusion graph in every RSG cycle, as described in Fig.~\ref{fig:fusiongraph}c. (b) If the 1-delay of these modules is replaced by a $k$-delay, then these modules can produce a larger fusion graph consisting of $k$ disconnected cuboids that each have 2D slices of size $n_x \times n_y$.}
\label{fig:l1modules}
\end{figure*}

Quantum computations with static matter-based qubits are usually described as quantum circuits that are executed by an array of physical qubits.
We refer to this paradigm as \textit{circuit-based quantum computation} (CBQC) to distinguish it from photonic FBQC. 
In fault-tolerant CBQC with topological codes such as surface codes, the operations that appear in the quantum circuit (see Fig.~\ref{fig:overview}a) are typically single-qubit measurements and two-qubit gates between qubits that are neighbors when arranged on a 2D grid. 
Therefore, these operations can be implemented by a 2D grid of static qubits (such as superconducting qubits) executing these gates one layer of operations after the other, as shown in Fig.~\ref{fig:overview}b.

In photonic FBQC, both the description of the computation and the hardware implementation are different than in CBQC (see Fig.~\ref{fig:overview}c-f). Instead of a circuit, the operations required to execute a specific computation are described by a \textit{fusion graph}. 
Instead of arrays of static qubits, these operations are executed by networks of modules consisting of resource-state generators, fusion devices and photonic delay lines, connected via waveguides and switches. 
In this section, we review the structure of simple cubic fusion graphs using 6-ring graph states as resource states. 
We show how such fusion graphs can be implemented by hardware modules. 
While we describe the fusion-graph structure in this section, the exact correspondence between this FBQC protocol and fault-tolerant quantum computations with surface codes is discussed in Sec.~\ref{sec:logic}.

\textbf{Fusion graphs.} 
In FBQC, a quantum computation corresponds to a set of fusion instructions applied to an arrangement of resource states, which is described by a \textit{fusion network}~\cite{FBQCpaper}.
For our purposes, a useful description of such a computation is a fusion graph, which can be used in the special case that each fusion operation involves at most two resource states.
The fusion graph contains vertices, edges connecting pairs of vertices, and half-edges connected to a single vertex. Here, each vertex corresponds to an $m$-qubit resource state, where $m$ is the number of edges and half-edges connected to the vertex. 
Each qubit in a fusion graph is either part of a single-qubit measurement (half-edge) or a two-qubit fusion (edge).
Throughout this article, the examples we present are all based on the repeated use of a single kind of resource state, making the fusion graph representation a natural fit.

We focus on the example of a specific resource state with $m=6$, the 6-ring. 
This is a six-qubit ring-shaped cluster state~\cite{Raussendorf2001}, so it can be thought of as the state that is obtained by initializing six $|+\rangle = (|0\rangle + |1\rangle)/\sqrt{2}$ states and performing controlled-$Z$ gates between pairs of neighboring qubits arranged on a ring.
Note that such states can be prepared using linear-optical components without the use of controlled-$Z$ gates~\cite{Browne2005}. 
The fusion graph describing fault-tolerant FBQC with 6-ring resource states is a particularly simple cubic lattice, as shown in Fig.~\ref{fig:fusiongraph}a. 
In the bulk of the fusion graph, each vertex has six neighbors, so each resource state is fused with six other resource states, one fusion per constituent qubit of the resource state.

For clarity, we will refer to the axes of the simple cubic lattice as $x$, $y$ and $z$. 
We will similarly label the qubits of the corresponding resource states as $x^+$, $y^+$, $z^+$, $x^-$, $y^-$ and $z^-$.
Fusions between $x^\pm$, $y^\pm$ and $z^\pm$ qubits are referred to as $x$, $y$ and $z$ fusions, respectively.
These fusions are type-II fusions~\cite{Browne2005}, which means that they correspond to Bell-basis measurements of pairs of photonic qubits. 
These measurements yield two bits of information: the outcomes of the $Z \otimes Z$ and $X \otimes X$ measurements of the fused photonic qubits, where $X$ and $Z$ are Pauli operators. 

\textbf{FBQC hardware components.} To execute the operations described by a fusion graph, a photonic fusion-based quantum computer uses the hardware components shown in Fig.~\ref{fig:overview}d. 
Resource-state generators (RSGs) are elementary hardware components that produce one resource state every $t_{\mathrm{RSG}}$, a unit of time denoted by \clock~in the figures and referred to as an \textit{RSG cycle}. 
The qubits produced by RSGs are encoded as photonic qubits, which can be any of several possible ways of representing a qubit in photonic modes.
These can be encoded using a single photon, e.g., via polarization, time-bin or dual-rail encoding, or multiple photons, e.g, using GKP bosonic encoding \cite{Gottesman2001}. 
Moreover, multiple such qubits can be combined to encoded photonic qubits~\cite{FBQCpaper} using a stabilizer code such as a Shor code.
Regardless of the encoding, photonic qubits are transported by waveguides to other components, namely $n$-delays, fusion devices and switches. 
$n$-delays are passive components that delay (store) photons for $n$ RSG cycles, i.e., a photonic qubit entering an $n$-delay will exit the delay after $n$ RSG cycles. 
$n$-delays can store $n$ photonic qubits, thereby acting as a fixed-time quantum memory. 
Fusion devices perform entangling fusion measurements of photon pairs that enter the device. 
Finally, switches are active components used to reroute photons on demand. 
Incoming photonic qubits can be routed to one of multiple outgoing waveguides. 
Switch settings can be adjusted in every RSG cycle, thereby deciding which operations are performed.

These hardware components are motivated by integrated silicon photonics. 
Possible implementations are dual-mode photons in pairs of silicon waveguides as qubits, electro-optical modulators as switches, and low-loss optical fiber as $n$-delays. 
Free-space delay lines can also be used as $n$-delays. 
Fusion devices can be constructed using beam splitters and single-photon detectors.
Furthermore, linear-optical components can be used to construct RSGs~\cite{Carolan2015, Vigliar2019}. 
Because sources~\cite{sourcespeed}, electronics, and electro-optical modulators~\cite{switchspeed1,switchspeed2} operate on time scales characterized by GHz clock rates, the relevant time scale for resource-state generation is $t_{\text{RSG}}\approx 1$~ns.
If transduction from static solid-state qubits to suitable photons (i.e., photons at a frequency for which low-loss delays are possible) is available, RSGs can also be matter-based, making the architecture presented in the following compatible with solid-state components. 
The possibility of such a hybrid solid-state/photonic architecture is discussed in further detail in Sec.~\ref{sec:hardware}.

\textbf{Interleaving coordinates.} A fusion-based quantum computer consists of a fixed number of RSGs, denoted by $n_{\mathrm{RSG}}$. 
Typically, the number of resource states (vertices) in a fusion graph is substantially larger than $n_{\mathrm{RSG}}$, so they need to be generated in multiple RSG cycles. 
A fusion graph for implementing a logical gate has no notion of time.\footnote{The fusion graph of an entire fault-tolerant computation will have some limited time ordering imposed at the level of logical gates, where feed-forward is required by the quantum algorithm, but this happens at a scale much larger than individual resource states and fusion measurements. } 
While the fusion graph determines which operations need to be performed, it does not dictate in which order its resource states are generated. 
We will focus on a particularly simple fusion graph consisting of a cubic lattice of size $r_x \times r_y \times r_z$. 
In this case, it is useful to partition such a graph into $r_z$ 2D slices of size $r_x \times r_y$, as shown in Fig.~\ref{fig:fusiongraph}b.

To implement a fusion graph using a network of $n_{\mathrm{RSG}}$ RSGs, we assign \textit{interleaving coordinates} $(g,t)$ to each vertex of the fusion graph, where the RSG coordinate $g$ specifies which RSG is to generate the resource state corresponding to the vertex, and the time coordinate $t$ specifies the RSG cycle in which it will be generated. 
All fusion-graph vertices must be assigned different interleaving coordinates with $g \leq n_{\mathrm{RSG}}$. 
A particularly simple assignment is shown in Fig.~\ref{fig:fusiongraph}c, where all vertices part of the same slice have different RSG coordinates $g$, but identical time coordinates $t$, with $t$ corresponding to the $z$ coordinate of the slice.

We refer to fusions between photonic qubits from resource states with identical time coordinates as \textit{instantaneous fusions}, and with different time coordinates as \textit{delayed fusions}.
This distinction can be relevant, as only instantaneous fusions may be realized without transitioning into delay elements.
Moreover, we refer to fusions between photonic qubits produced by different RSGs as \textit{networked fusions}, and by the same RSG as \textit{local fusions}. 
In the coordinate assignment of Fig.~\ref{fig:fusiongraph}c, all $x$ fusions and $y$ fusions are instantaneous networked fusions, whereas $z$ fusions are delayed local fusions with a time delay of $t_{\text{RSG}}$.

A network of hardware modules capable of implementing a fusion graph with such a coordinate assignment is shown in Fig.~\ref{fig:l1modules}a. It is an array of $n_x \times n_y$ modules each consisting of one RSG, three fusion devices and a 1-delay. Each module has up to four neighboring modules, which we refer to as the north, east, south and west neighbors. The fusion device in the lower right corner of each module is responsible for $y$ fusions. These are networked fusions between a photon produced in this module and a photon produced in the neighboring module to the south. Similarly, the middle fusion device is responsible for $x$ fusions between qubits from this module and the neighboring module to the east. Finally, the top fusion device is responsible for $z$ fusions, which are delayed local fusions. $z^+$-qubits need to be fused with $z^-$-qubits that will be generated in the subsequent RSG cycle. Since their fusion partners do not exist yet when they are produced, $z^+$-qubits go through a 1-delay before entering the fusion device. Correspondingly, $z^-$-qubits are fused with qubits produced in the previous RSG cycle, i.e., qubits from the previous slice of the cubic fusion graph. Such a network of modules can implement a fusion graph with 2D slices of size $n_x \times n_y$, using 1 RSG cycle per slice. 
We refer to these modules as \textit{trivial interleaving modules}, as they mimic a 2D array of static qubits sequentially executing gates and measurements, where each fusion-graph slice can be seen as analogous to a layer of operations.

\begin{figure*}[t]
\centering
\includegraphics[width=0.95\linewidth]{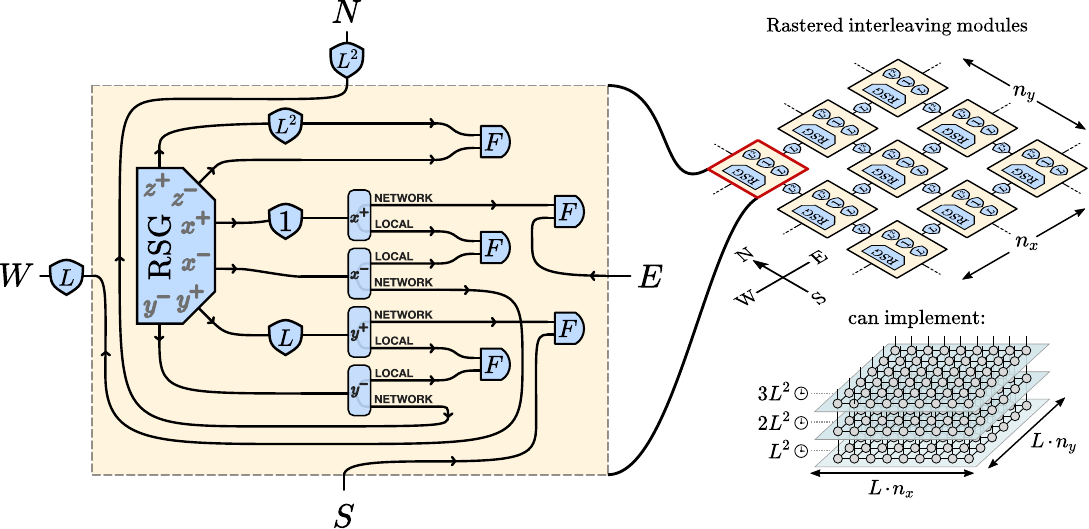}
\caption{Interleaving modules that are compatible with the interleaving coordinate assignment shown in Fig.~\ref{fig:fusiongraph}d. An array of $n_x \times n_y$ modules can produce a fusion graph with 2D slices of size $L \cdot n_x \times L \cdot n_y$, where each slice is produced in $L^2$ RSG cycles.}
\label{fig:bulkmodule}
\end{figure*}

\textbf{Interleaving.} In the context of computing, interleaving refers to the task of allocating the same memory or processing power to different tasks in succession. 
In the specific context of FBQC, interleaving refers to the allocation of the same RSG for the successive production of different fusion-graph resource states. 
Figures \ref{fig:fusiongraph}c and \ref{fig:l1modules}a provide a trivial version thereof, where a 2D arrangement of RSGs is used to produce one slice of resource states after the other.
We are particularly interested in allocation schemes which allow the production of larger fusion graphs. 
Such schemes become available by using longer delays.  

A particularly simple example demonstrating that such delays can effectively increase the size of the fusion graph that can be implemented is shown in Fig.~\ref{fig:l1modules}b. 
This is a network of modules that is identical to the one discussed previously, except that the 1-delay is replaced with a $k$-delay. 
Such a network produces $k$ copies of disconnected fusion graphs with 2D slices of size $n_x \times n_y$. 
Each slice is now effectively produced in $k$ RSG cycles instead of 1 RSG cycle. 
By extending the delay in the module from a $1$-delay to $k$-delay we effectively multiply the number of active qubits by $k$, which manifests in the network creating $k$ identical copies of the original fusion graph which are ``interleaved'' in time. 
As an example, if the original hardware (with a 1-delay) could be used to create a single logical qubit, then this modified system can be used to create $k$ logical qubits. 
We refer to this approach as \emph{layered interleaving}.

\textbf{Rastered interleaving modules.}
Alternatively, we can consider a different time ordering of operations to use a single RSG to create an entire patch of a fusion graph one resource state at a time. 
To this end, we change the assignment of interleaving coordinates, as shown in Fig.~\ref{fig:fusiongraph}d. Each slice with a $z$ coordinate $z_s$ is partitioned into squares of size $L \times L$, where we refer to $L$ as the \textit{rastering length}. In each such square, all resource states are assigned the same RSG coordinate $g$ and a \textit{time bin} $b \in \{1,\dots L^2 \}$ in ascending order from left to right and top to bottom. The time coordinates $t$ of each resource state are then chosen as $L^2 z_s+b$. 
In this coordinate assignment, all fusions are delayed fusions. Fusions between resource states within a square are local fusions, whereas fusions between different squares are networked fusions. We refer to this approach as \emph{rastered interleaving}. 
The approach of using a single physical resource to create a higher-dimensional entanglement structure has been considered before in other physical contexts~\cite{lindner2009proposal}, and making use of the regular cubic 3D structure was also recently proposed in Ref.~\cite{Wan2020}.

A network of modules capable of implementing such a coordinate assignment is shown in Fig.~\ref{fig:bulkmodule}. We refer to these modules as \textit{rastered interleaving modules}. 
They consist of one RSG, five fusion devices, four switches and delays of length $1$, $L$ and $L^2$. 
Modules are connected to neighboring modules via $L$-delays to the west and via $L^2$-delays to the north. 
As in the previous case, $z$ fusions are local fusions between two $z$ slices. 
However, their time difference has increased from $1$ to $L^2$ RSG cycles, so the $z^+$ qubit enters an $L^2$-delay.

There are now two types of $x$ fusions. 
Local $x$ fusions have a time difference of 1 RSG cycle, whereas networked $x$ fusions are separated by $L-1$ cycles. 
For this reason, the $x^+$ and $x^-$-qubit both enter a switch to either route them to a local fusion device or a networked fusion device. 
When an $x^-$-qubit is sent to the neighboring module to the west, it is delayed by $L$ RSG cycles. 
Since $x^+$-qubits are always delayed by 1 RSG cycle before entering the switch, this leads to a time difference of $L-1$ cycles for networked fusions. 
Similarly, local $y$ fusions have a difference of $L$ cycles, whereas networked $y$ fusions have a difference of $L^2-L$ cycles.

\begin{figure}[b]
\centering
\includegraphics[width=\linewidth]{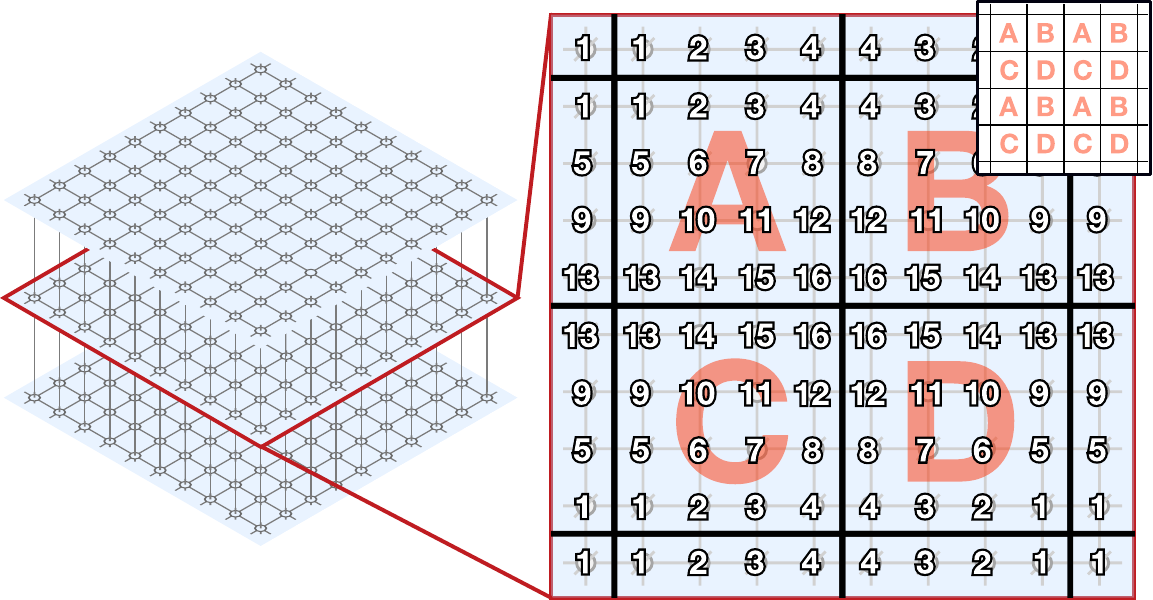}
\caption{Alternative assignment of interleaving coordinates in which all networked fusions are instantaneous fusions.}
\label{fig:abcdorder}
\end{figure}

Networks of $n_x \times n_y$ interleaving modules can produce slices of size $L \cdot n_x \times L \cdot n_y$ using $L^2$ RSG cycles per slice. 
As we show in Sec.~\ref{sec:logic}, this corresponds to a linear space-time trade-off: an increase in the number of logical qubits by a factor of $L^2$ and a decrease in computational speed by the same factor, compared to modules that only use 1-delays. Conversely, interleaving modules reduce the number of RSGs required to encode a fixed number of logical qubits by a factor of $L^2$. We define the ratio of $n_{\mathrm{slice}}$ to $n_{\mathrm{RSG}}$ as the \textit{interleaving ratio}, where $n_{\mathrm{slice}}$ is the size of fusion-graph slices (measured in the number of resource states) that can be produced by a network of modules, and $n_{\mathrm{RSG}}$ is the total number of RSGs in the network. For rastered interleaving modules, the interleaving ratio is $L^2$.

Note that it is possible to remove the $L$- and $L^2$-delays between neighboring interleaving modules by slightly altering the assignment of interleaving coordinates. If the $L\times L$ squares are labeled A, B, C and D in the alternating fashion shown in Fig.~\ref{fig:abcdorder}, then all networked fusions become instantaneous fusions. For B segments, time-bin labels increase right to left. For C segments, they increase bottom to top. For D segments, they increase both right to left and bottom to top. In this case, interleaving modules come in four types (A, B, C and D) which feature the same types of delays, but differ slightly in the arrangement of their hardware components.

\section{Universal Interleaving Modules}
\label{sec:logic}

In this section, we review a prescription for fault-tolerant fusion-based quantum computations (FBQC) with surface codes and 6-ring fusion graphs. 
We demonstrate that, with small modifications, the networks of interleaving modules introduced in the previous section are universal fault-tolerant quantum computers. 
We show explicit constructions for the implementation of a universal set of logical operations using lattice surgery.
The construction is in some cases analogous to that of measurement-based quantum computing (MBQC)~\cite{Raussendorf_2007}. 
However, while the operations in MBQC are the preparation of a large cluster state and single-qubit measurements on cluster-state qubits, the operations in FBQC are the repeated generation of small identical resource states and entangling measurements between pairs of resource states.

\subsection{From surface codes to fusion graphs}

\begin{figure*}[t]
\centering
\includegraphics[width=0.96\linewidth]{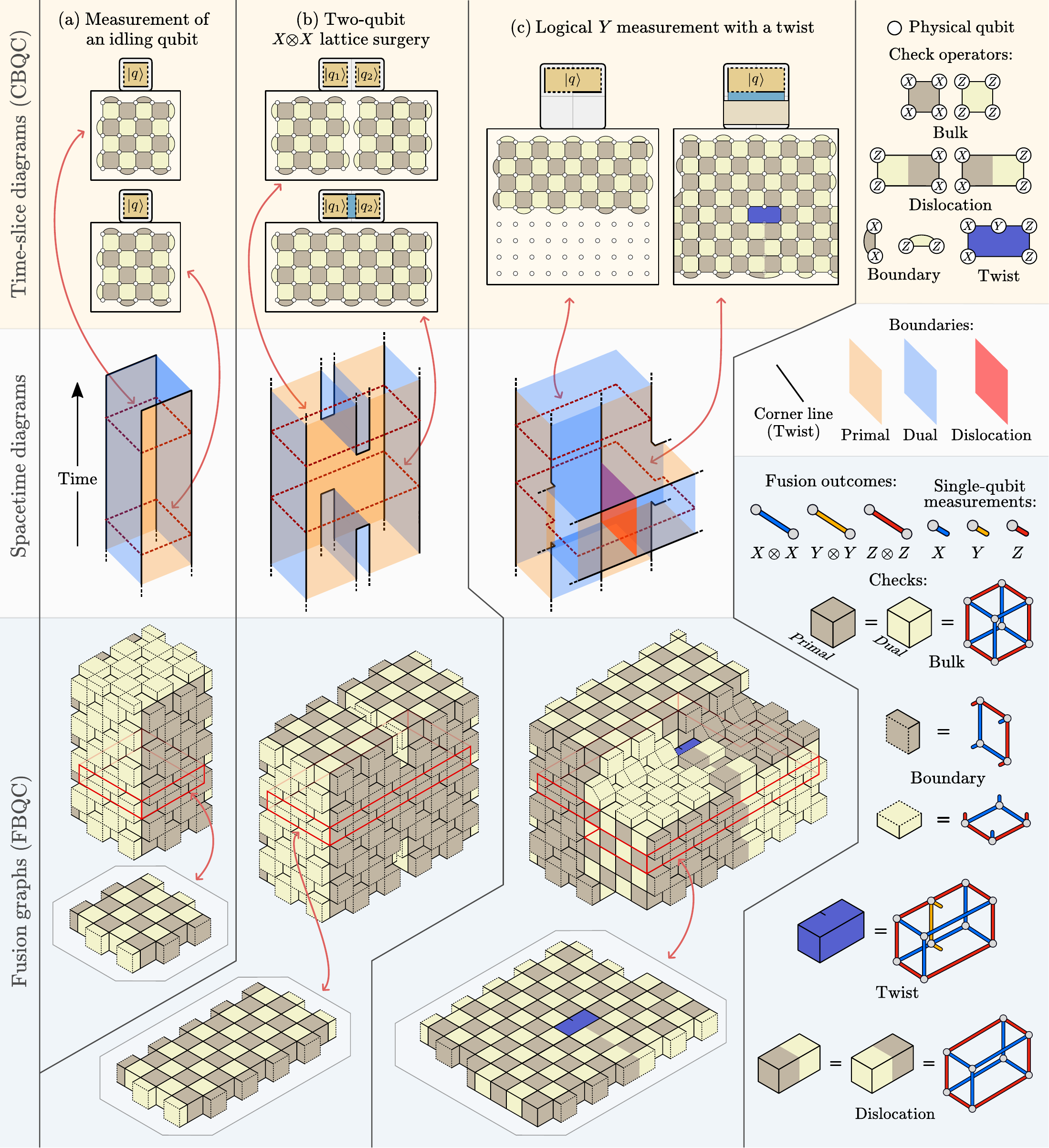}
\caption{Three examples showing the translation between surface-code spacetime diagrams, time-slice diagrams and 6-ring fusion graphs. Time slices show a 2D array of physical qubits and the check operators that are measured during a specific time step. Spacetime diagrams are 3D diagrams with time increasing in the upward direction. Solid black lines and orange and blue surfaces trace the trajectory of corners and primal and dual boundaries through spacetime. Fusion graphs are generated by filling the bulk of the spacetime diagram with primal and dual bulk-check cubes in a 3D checkerboard pattern. 
Boundaries are decorated with primal or dual half-cubes, depending on the type of the boundary. 
Each row of the figure is discussed in more detail throughout corresponding subsections of Sec.~\ref{sec:logic}.}
\label{fig:logicoverview}
\end{figure*}

We first review spacetime diagrams of surface codes in the context of CBQC based on 2D arrays of physical qubits. When performing fault-tolerant quantum computations with surface codes, an entire quantum computation can be described as a sequence of \textit{check-operator measurements}, i.e., measurements of operators that implement certain logical operations and enable the detection and correction of errors. Quantum computations are performed in time steps (or \textit{code cycles}). In each time step, a certain set of check operators defined on the 2D array of physical qubits is measured. Figure~\ref{fig:logicoverview} shows two commonly used descriptions of protocols with surface codes: time slices and spacetime diagrams.

\textbf{Time slices.} In CBQC, a time-slice diagram specifies all check operators that are measured in a certain time step. Here, each circle corresponds to a physical qubit and each face corresponds to a check operator. For instance, one possible way of encoding a single logical qubit with surface-code patches is via a square patch, as shown in the time slices of Fig.~\ref{fig:logicoverview}a. A square patch consists of $d \times d$ physical qubits, where $d$ is the code distance and $d=5$ in the example drawn in the figure. The check operators of such a patch are four-qubit operators $X^{\otimes 4}$ and $Z^{\otimes 4}$ in the bulk and two-qubit operators $X^{\otimes 2}$ and $Z^{\otimes 2}$ at the boundary, where $X$, $Y$ and $Z$ are the Pauli operators of physical qubits. Time slices can also be drawn using a simplified representation that ignores the microscopic details of physical qubits and check operators. In the simplified representation of the time slices in Fig.~\ref{fig:logicoverview}a, square patches encoding the qubit $|q\rangle$ are drawn as squares labeled $|q\rangle$.

A four-corner patch has four boundaries, which are drawn as solid and dashed edges in the simplified representation. These boundaries can be thought of as encoding the logical Pauli operators of the logical qubit, where solid and dashed edges correspond to logical $Z$ and $X$ operators, respectively. They are also referred to as primal and dual, rough and smooth, or $Z$ and $X$ boundaries.

\begin{figure*}[t]
\centering
\includegraphics[width=\linewidth]{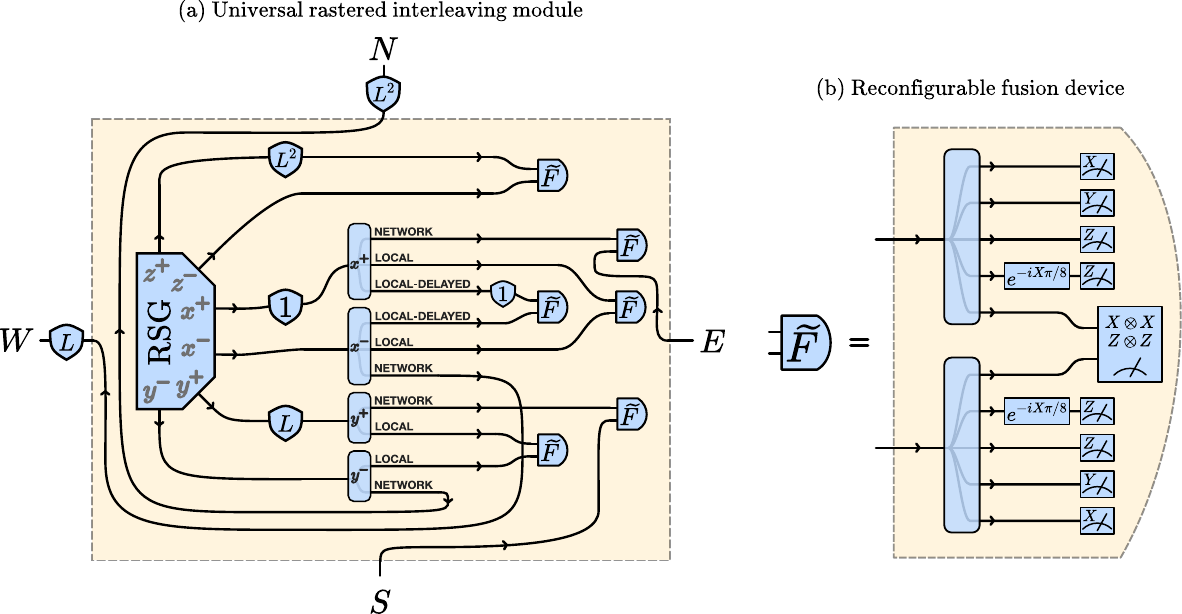}
\caption{Interleaving modules capable of universal quantum computation. These are almost identical to the modules in Fig.~\ref{fig:bulkmodule}, except that fusion devices are replaced with reconfigurable fusion devices and that the additional \textit{local-delayed} fusion can be used to implement lattice dislocations and twist defects in surface codes.}
\label{fig:logicmodule}
\end{figure*}

\textbf{Spacetime diagrams.} While it is possible to describe an entire quantum computation or protocol as a sequence of time slices, it is often more convenient to draw multiple 2D time slices in a 3D diagram, a spacetime diagram. In such a diagram, time can be thought of as increasing in the upward direction, whereas the other two dimensions represent space. The solid black lines in a spacetime diagram trace the trajectory of patch corners through spacetime. Similarly, orange and blue surfaces track primal and dual boundaries through spacetime. Time slices then correspond to two-dimensional spacelike cuts through the spacetime diagram. Corner lines and boundaries can also move in spacelike directions, which typically happens for measurements of logical qubits. For instance, the example in Fig.~\ref{fig:logicoverview}a describes a logical qubit $|q\rangle$ that idles for a while, until it is measured in the $Z$ basis, as indicated by the corner lines and the dual boundary capping off the spacetime diagram in time. The final time slice of this protocol (not drawn) would correspond to all physical qubits being measured in the $Z$ basis.

The remaining two protocols in Fig.~\ref{fig:logicoverview}b and c correspond to a logical two-qubit $X \otimes X$ measurement via lattice surgery~\cite{Horsman2012} and a logical qubit encoded in a rectangular patch contributing to a multi-qubit Pauli measurement with its $Y$ operator~\cite{Litinski2019}. In the context of this paper, it is not important to understand what these logical operations are, or why the shown spacetime diagrams correspond to these operations. These examples are chosen because they involve all features that are typically encountered in surface-code protocols: bulk checks, primal and dual boundaries, twist defects and lattice dislocations. Twist defects~\cite{Bombin2010} are found in time slices whenever a corner line is not located at the boundary of a spacetime diagram, but passes through the bulk. They are accompanied by a lattice dislocation that either extends to a boundary or a second twist defect.

On a final note, the spacetime diagrams in Fig.~\ref{fig:logicoverview} do not directly show how many time slices they correspond to. Typically (but not generally), whenever the spatial configuration changes in a spacetime diagram, this new configuration corresponds to $d$ time slices. For instance, the configuration of two merged patches in Fig.~\ref{fig:logicoverview}b would correspond to $d=5$ time slices, of which only one is drawn.

\textbf{Surface-code checks in fusion graphs.} Before we outline how spacetime diagrams are translated into fusion graphs, we first introduce the notation for 6-ring fusion graphs used in Fig.~\ref{fig:logicoverview}. As in Fig.~\ref{fig:fusiongraph}a, these fusion graphs describe a cubic lattice of resource states, but now contain additional information with colors assigned to certain cubic volumes within the fusion graph. 
Such a fusion graph can not only be used to read off which pairs of qubits need to be fused, but also indicates how the classical fusion outcomes can be assembled to surface-code check measurement outcomes that can be fed into a decoder. As shown in Fig.~\ref{fig:logicoverview}, surface-code bulk checks correspond to products of 12 fusion-measurement outcomes involving eight resource states arranged in a cube. Out of these 12 outcomes, six are $X \otimes X$ outcomes and six are $Z \otimes Z$ outcomes. (As a reminder, each two-qubit fusion produces two measurement outcomes: an $X \otimes X$ and a $Z \otimes Z$ outcome.) In Fig.~\ref{fig:logicoverview}, $X \otimes X$ and $Z \otimes Z$ outcomes are drawn as blue and red edges, respectively. Both primal and dual checks correspond to the same hardware operations and combinations of outcomes, and only differ in how the information is used during decoding. The reason why these cubes correspond to check operators is that the product of 12 fusion-measurement outcomes is equivalent to a product of eight resource-state stabilizers of the eight 6-ring cluster states corresponding to the corners of the cube, as explained in Ref.~\cite{FBQCpaper}. Boundary checks correspond to half-cubes that not only involve fusion outcomes, but also single-qubit measurement outcomes. These are drawn as half-edges corresponding to the measurement of one of the qubits of the resource state located at the vertex, namely the qubit corresponding to the direction of the half-edge. Twists involve $Y \otimes Y$ fusion outcomes. These do not require additional hardware operations, but are obtained by multiplying $X \otimes X$ and $Z \otimes Z$ outcomes.

\textbf{From spacetime diagrams to fusion graphs.} The cubic fusion graph corresponding to a specific spacetime diagram can be obtained using a remarkably simple procedure. The bulk of the fusion graph is filled with primal and dual bulk-check cubes in a 3D checkerboard pattern, and the primal and dual boundaries are decorated with primal or dual half-cubes. 
If twists or lattice dislocations are present, they are added using the corresponding checks. 
Note that slices of the fusion graph highlighted in Fig.~\ref{fig:logicoverview} mimic the pattern of the corresponding CBQC time slices. 
The number of cubes in the fusion graph determines the code distance, where time slices of distance-$d$ square patches involve $d^2$ resource states.

It is unsurprising that fusion graphs and time slices look similar, since they are obtained from similar considerations. 
For time slices in CBQC, each physical qubit in the bulk should be part of two dual and two primal checks, such that both $X$ and $Z$ Pauli errors can be detected and corrected. At a boundary, each qubit is part of two dual and one primal, or two primal and one dual check, depending on the type of the boundary. Two distinct equal-type boundaries are separated by at least $d$ physical qubits, so that it takes at least $d$ Pauli errors between time slices to generate an undetectable error pattern. Similarly, each fusion-measurement outcome in the bulk of the fusion graph is part of two primal and two dual checks, so that $X$ and $Z$ errors affecting the qubits that are part of the fusion can be detected and corrected. At boundaries, fusions are part of two primal (or dual) checks and one dual (or primal) check. Distinct equal-type boundaries are separated by at least $d$ fusions. 
The fault-tolerant FBQC protocols using the 6-ring resource state in a way corresponding to CBQC with surface codes are described in more detail in Ref.~\cite{LogicBlocksPaper}.

\subsection{Universal interleaving modules}

\begin{figure}[b]
\centering
\includegraphics[width=0.9\linewidth]{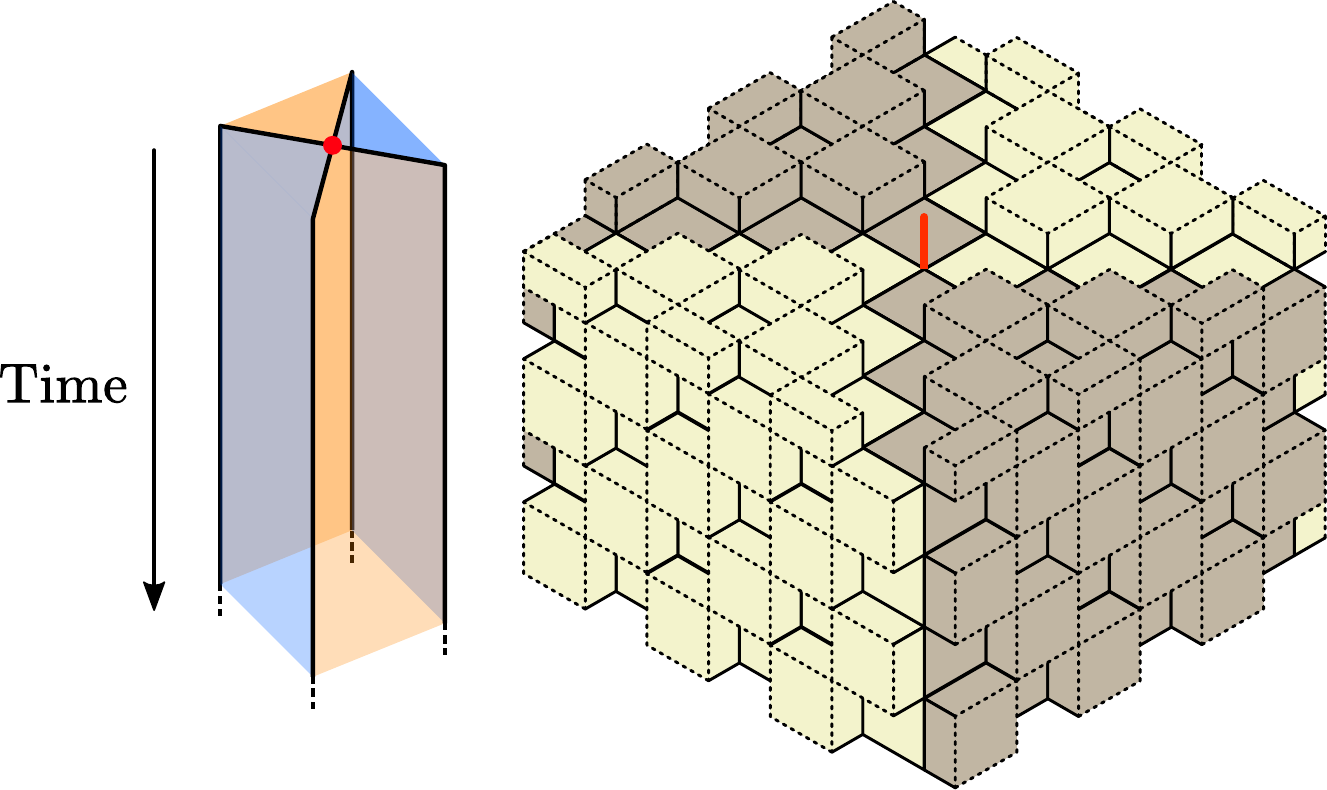}
\caption{Example of a state-injection protocol~\cite{Lodyga2015, Brown2020} for the preparation of a faulty magic state. The red half-edge corresponds to an $e^{-iX\pi/8}$ rotation (a $T$ gate) followed by a $Z$ measurement. Note that, in contrast to other spacetime diagrams in this paper, time increases in the downward direction, so that the relevant operation is visible.}
\label{fig:injection}
\end{figure}

Since the bulk of 6-ring fusion graphs corresponding to surface-code spacetime diagrams is still a regular cubic lattice, the interleaving modules of Fig.~\ref{fig:bulkmodule} require only small modifications. Interleaving modules capable of universal quantum computation are shown in Fig.~\ref{fig:logicmodule}. 
The main modification is the replacement of all fusion devices with \textit{reconfigurable} fusion devices. Here, the two incoming photons enter a switch that routes the qubits either to a two-qubit fusion measurement or a single-qubit measurement. 
Universal quantum computers require at least one non-Clifford operation~\cite{Gottesman1999}. 
Therefore, the switch includes the option to perform an $e^{-iX\pi/8}$ rotation (a $T$ gate) followed by a $Z$ measurement. 
In the context of surface codes, such operations are required for state injection of magic states, i.e., the non-fault-tolerant preparation of $T$-gate resource states. An example of a state-injection protocol~\cite{Lodyga2015, Brown2020} is shown in Fig.~\ref{fig:injection}, where the red half-edge corresponds to such a $T$ gate followed by a measurement.

The second modification is the addition of another switching option and fusion device for $x$ fusions. 
Assuming that dislocations are always located along the $x$ direction and that the horizontal extent of dislocations is never located at the boundary between two neighboring modules, dislocations then correspond to delayed local fusions with a time difference of two RSG cycles. The additional switching option in Fig.~\ref{fig:logicmodule} therefore delays the $x^+$-qubit by one additional RSG cycle.

\textbf{Universal operations and full quantum computations.}
The logical operations shown in Figs.~\ref{fig:logicoverview} and \ref{fig:injection}, and small variations thereof, are sufficient for universal quantum computation.
One possible set of universal logical operations compatible with surface codes and lattice surgery consists of the preparation of magic states and the measurement of arbitrary multi-qubit Pauli operators.
While the former is shown in Fig.~\ref{fig:injection}, the latter can be constructed as a string of logical operations similar to the one shown in Fig.~\ref{fig:logicoverview}c and is explained in greater detail in Ref.~\cite{Litinski2019}.

Therefore, the logical operations discussed in this section can be directly used as a prescription to translate any arbitrary quantum computation into sequences of switch settings for a network of universal interleaving modules, such as those shown in Fig.~\ref{fig:logicmodule}.
A full quantum computation corresponds to a sequence of switch settings in such a network.
When executing logical operations using lattice surgery, the prescription outlined in this section can be used to determine the switch settings for every RSG cycle of the computation.

\textbf{Beyond 6-ring fusion networks.} It is worth commenting that we have focused entirely on the specific example of the 6-ring fusion network, but the principles of interleaving extend beyond this one example.
When generating the bulk of the 3-dimensional fusion graph, we are exploiting its translational invariance to repeatedly reuse identical resource-state generator modules.
Furthermore, the regularity of the structure implies that each resource state in the bulk can repeatedly reuse the same delay lines.
While the fusion graph in our examples is a simple cubic lattice, in general, such a translational invariance can be exploited for any fusion graph that is an arbitrary-dimensional Bravais lattice.
To implement logical operations, the translational invariance needs to be broken in order to introduce topological features.
This requires switching, either of the measurement basis of certain fusions, or of the locations to which qubits are routed.

\section{Benefits of Interleaving in Architectural Design}
\label{sec:architecture}

In the previous sections, we have described a scalable modular architecture for fault-tolerant quantum computing that uses optical fiber as quantum memory and as macroscopic optical interconnects between interleaving modules. In this section, we discuss the advantages of such an architecture. 
We first study realistically achievable fiber-delay lengths and interleaving ratios by presenting numerical results for the loss threshold.
Next, we focus on the linear space-time trade-off provided by interleaving and the potential reduction in the cost of logical operations that can be gained by taking advantage of the flexible connectivity between modules. Finally, we 
argue that interleaving is not only compatible with purely photonic architectures, but can also provide an alternative approach to scaling up matter-based devices, if these can be used for resource-state generation.

\begin{figure}[t]
 \centering
    \includegraphics[width=\linewidth]{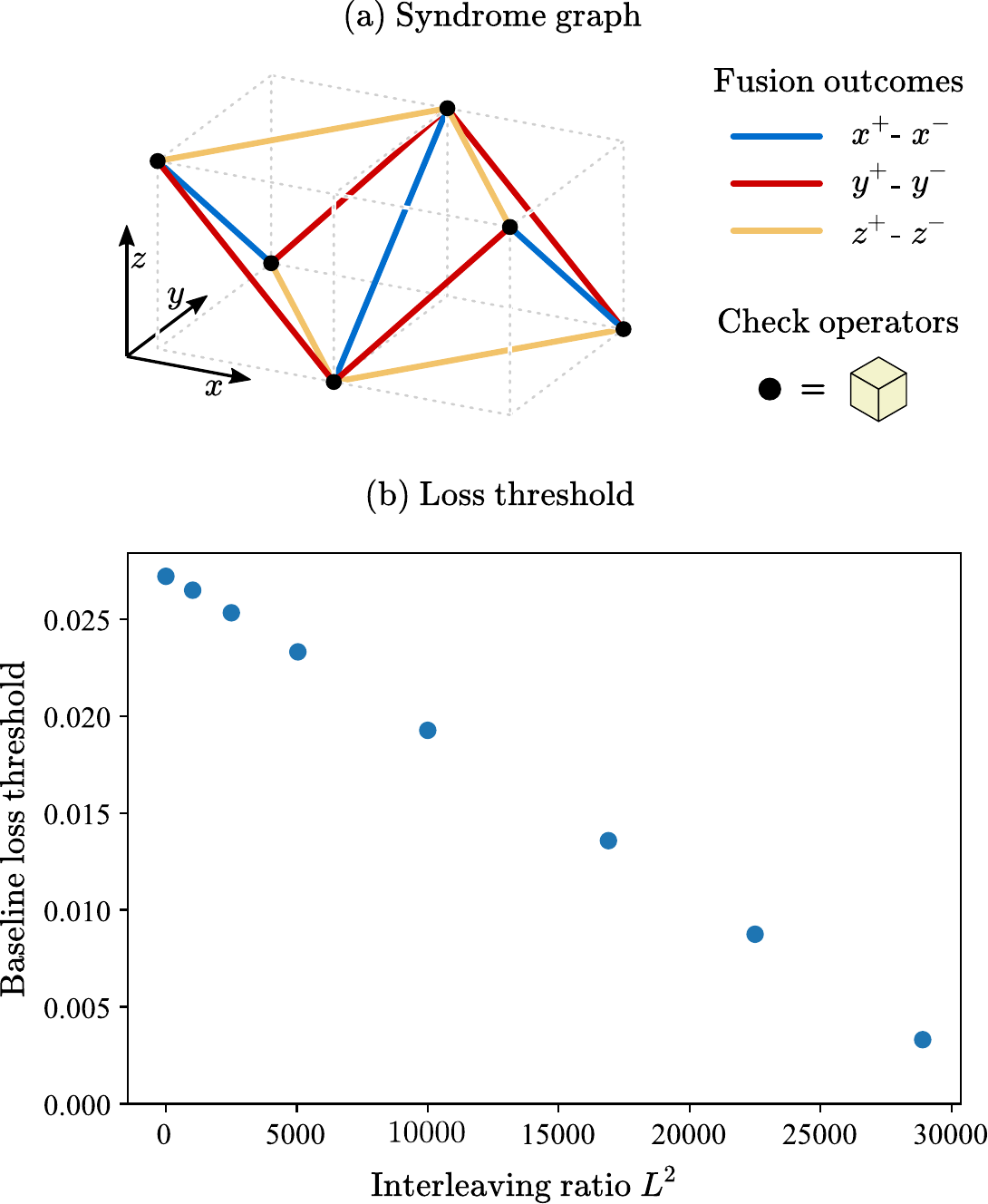}
    \caption{Simulation of the effect of delay loss. 
    (a) The syndrome graph for the simple cubic 6-ring fusion network where edges corresponding to fusions in different orientations are distinguished by their color. 
    Each fusion contributes two edges to the syndrome graph, one to the primal and one to the dual. 
    The primal and dual syndrome graphs are identical. 
    (b) The results of numerical simulations of the syndrome graph in (a), where each photonic qubit (and fusion) is encoded using a 4-qubit Shor error-detecting code over dual-mode photonic qubits and loss rates for each edge are set according to an interleaving scheme with varying interleaving ratios $L^2$. 
    The plot shows the decrease in baseline loss threshold with increasing interleaving ratio, where the baseline loss is the non-fiber loss rate of each resource-state photon. In other words, the baseline loss threshold refers to how much photon loss may be tolerated during resource-state generation and switching before entering the delays and in the switching and measurement after the delays.}
    \label{fig:loss_threshold}
\end{figure}

\subsection{Increased number of logical qubits due to high delay-line loss tolerance}
\label{sec:numerics}

To make use of the methods we have described, we need to assess the impact of the additional delay loss on the fault-tolerance threshold. 
Given that photon loss is the primary concern with long delays, a useful insight into the behavior of an interleaved system can be obtained by studying a simple model of photon loss that accounts for the additional loss of the fiber on different qubits of the resource state. 

We consider the example of using the rastered-interleaving scheme to generate the photonic fusion network introduced in Ref.~\cite{FBQCpaper}. 
This scheme is based on the 6-ring network, and uses encoded fusion operations based on the 4-qubit Shor code. 
We describe more details of this scheme in Appendix~\ref{app:simulation}. 
With trivial interleaving (i.e., an interleaving ratio of $L^2=1$), and with an error model where each photon experiences a loss of the same rate, $p$, the scheme has a threshold of $2.7\%$ to photon loss.  
To study the behavior when we add interleaving, we consider an error model where each photon of a resource state suffers both a \emph{baseline loss} with a rate of $p_{\rm baseline}$ which is due to the process of the creation, routing and measurement of the state, and an additional loss $p_{\rm delay} = 1-(1-p_{\text{\clock}})^N$ that depends on the length of the delay line in the interleaving module. $N$ indicates the number of time bins in the delay, and $p_{\text{\clock}}$ is the loss per time bin. When longer interleaving delays are used, the tolerance to the baseline loss will decrease. 

\begin{table}[t]
\begin{tabular}{p{0.9cm}p{1.2cm}p{1.5cm}p{2.6cm}p{1.6cm}}
$L$   & $L^2$ & Longest fiber & Largest delay loss & Threshold $p_{\mathrm{baseline}}$ \\
\hline
1   & 1        &       0.2 m         &       0.0009\%    &   2.7\% \\
32  & 1024      &      205 m        &      0.94\%       &    2.6\%              \\
71  & 5041       &     1008 m      &         4.5\%      &     2.3\%            \\
100 & 10000     &       2000 m           &    8.8\%       &     1.9\%        \\
\end{tabular}
\caption{Numerical results for the decrease in baseline loss threshold with increasing rastering length $L$, and consequently increasing interleaving ratio $L^2$, length of longest fiber and delay loss due to the longest fiber.}
\label{tab:results}
\end{table}

In order to simulate this system, it is helpful to define the syndrome graph, which is shown in Fig.~\ref{fig:loss_threshold}a. By using the syndrome graph, we can use standard Monte Carlo sampling and decoding methods to evaluate the loss threshold. In this representation, each vertex represents a parity check, and each edge corresponds to a fusion-measurement outcome. 
The parity checks at the vertices are evaluated by combining the fusion-measurement outcomes from each of the incident edges, as described in Ref.~\cite{FBQCpaper}. 
Note that the cubic structure shown by the dotted lines is the \emph{dual} of the fusion graph, where each cube corresponds to one resource state.  There are 6 edges on the faces of each cube, representing one measurement outcome from each of the 6 fusions on the resource state. Each fusion produces two measurement outcomes, and the remaining 6 outcomes from each resource state form a separate, locally disconnected, dual syndrome graph. Here the primal and dual syndrome graphs are identical. The figure shows only a portion of the syndrome graph, and eight such cubes combine to make a repeating unit cell. Each edge (fusion measurement) has an associated probability of erasure. The interleaving module introduces direction-dependent loss on the resource state, and this leads to three distinct types of edge in the graph, each of which has a different probability of erasure.
We point out that the syndrome graph in Fig.~\ref{fig:loss_threshold}a is drawn in a different orientation compared to Ref.~\cite{FBQCpaper}, which is explained in greater detail in Appendix~\ref{app:simulation}.

To simulate the behavior of this system, we model loss on the resource state as random and independent loss applied after ideal state preparation. Each qubit $i$ of the state experiences a loss 
$$p_i = 1- (1-p_{\mathrm{baseline}})(1-p_{\text{\clock}})^{N_i} \, ,$$ 
where $p_{\mathrm{baseline}}$ is the baseline loss of each qubit that comes from state preparation, $p_{\text{\clock}}$ is the additional loss experienced for each clock cycle that the qubit spends in a fiber delay line, and $N_i$ is the number of RSG cycles that qubit $i$ spends in fiber. If the rastering length is $L$ and the interleaving ratio is $L^2$, then $N_{1-6} = \{0,0,0,1,L,L^2\}$. 
Note that $L$ is independent of the code distance $d$ of a logical qubit. As the code distance of the logical qubit is increased, more patches of size $L \times L$ are used to create it. The physical fusions in this architecture may use boosting via an ancillary Bell pair. 
All photons in these ancillary states suffer a loss $p_{\rm{baseline}}$, since they do not need to be delayed. Our goal is to understand how the addition of interleaving reduces the system's tolerance to baseline loss. We consider the following rastering lengths: $L=1$ (the case of trivial interleaving), and $L=32,50,71,100,130,150,170$.  
We assume that all delays are implemented with optical fiber with a loss of 0.2 dB/km (4.5 \%/km), and consider time bins of 1 ns duration. 
With a speed of light in fiber of 0.2 m/ns, this results in $p_{\text{\clock}}=$~0.04~mdB. 

We perform numerical threshold simulations for a 3D block with periodic boundary conditions in each dimension, and with code distances $d=$~{12, 16, 20}. For each rastering length, we sweep the value of $p_{\mathrm{baseline}}$ to identify a threshold with at least 15000 Monte Carlo samples to compute a logical error rate for each parameter combination. We use the union find decoder~\cite{Delfosse2020, unionfind}, which is optimal given our loss-only model. Additional details about the error model and simulations are provided in Appendix~\ref{app:simulation}. The results of these simulations are shown in Fig.~\ref{fig:loss_threshold}. A summary of this analysis is shown in Tab.~\ref{tab:results}.

These results demonstrate that very long delays and the corresponding  high loss in the fiber have a remarkably low impact on the tolerance to other errors. With trivial interleaving ($L=1$), the loss threshold is 2.7\% per photon. 
With an interleaving ratio of $L^2=1024$, the baseline threshold is reduced to 2.6\%. Even with $L^2 = 5041$, which requires a 1 km delay line, the baseline threshold is reduced only to 2.3\%. By pushing the delay length up to 2 km, we start to see a noticeable reduction in the threshold. The baseline loss threshold is now reduced to $1.9\%$. Note that the largest delay loss in Tab.~\ref{tab:results} can be higher than the baseline loss threshold, because it only affects $z^+$-photons, as only these photons enter an $L^2$-delay.

\begin{figure*}[t]
\centering
\includegraphics[width=\linewidth]{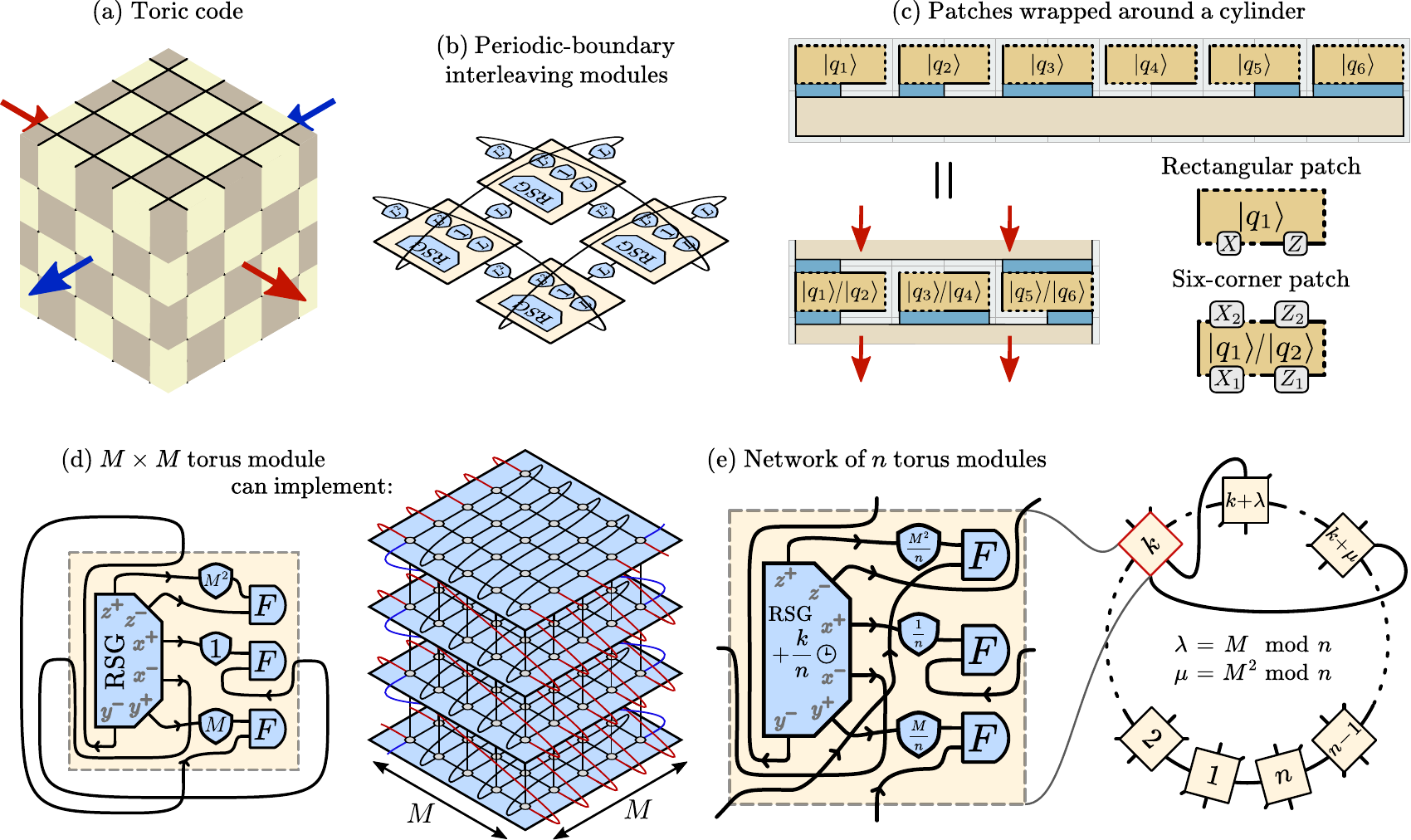}
\caption{Advantages of periodic boundary conditions. (a) Toric codes are surface codes with periodic boundary conditions and encode twice as many logical qubits using the same footprint. (b) With interleaving modules, periodic boundary conditions are easily achieved by connecting modules appropriately. (c) For planar surface codes, periodic boundary conditions in the vertical direction effectively reduce the footprint by a factor of two. (d) Alternatively, an $M \times M$ torus can be generated by a single module without any switching using a 1-, $M$- and $M^2$-delay. 
(e) To reduce the size of the maximum delay, $n$ modules with staggered RSGs can be connected to a network.
Note that if $M$ and $n$ are chosen such that $M= \pm 1 \mod n$, the interleaving modules and connections can be arranged geometrically locally in one dimension.}
\label{fig:periodic}
\end{figure*}

\subsection{Computational volume reductions due to linear space-time trade-offs}

For a fixed number of hardware components, the use of longer delays in interleaving increases the number of logical qubits that can be encoded. However, this comes at the cost of slower logical operations. In this sense, interleaving is a linear space-time trade-off: the number of qubits is increased by a factor of $L^2$, but the speed of logical operations is decreased by a factor of $L^2$. If we consider a device with a fixed number of RSGs and a specific quantum computation that we want to execute, some minimum interleaving ratio $L^2_{\rm min}$ is required to run this computation, as a minimum number of logical qubits need to be encoded. At the same time, the interleaving ratio cannot exceed $L^2_{\rm max}$, which is the interleaving ratio at which the loss threshold of the error-correcting code is reached. If $L^2_{\rm min} > L^2_{\rm max}$, then the quantum computation cannot be executed unless more hardware modules are added to reduce $L^2_{\rm min}$. Assuming that $L^2_{\rm min} < L^2_{\rm max}$, is it then advantageous to increase the interleaving ratio beyond $L^2_{\rm min}$?

Naively, the answer is no, as interleaving does not affect the volume of a computation. The fusion graph of a specific quantum computation will contain a certain number of resource states, which we refer to as the \textit{computational volume} $V_{\rm comp}$. The time $t_{\rm comp}$ to execute this computation then only depends on the volume, the number of resource-state generators $n_{\rm RSG}$ and the time to produce a resource state $t_{\rm RSG}$:
\begin{equation}\label{eq:rate_equation_0}
   t_{\rm comp} = \frac{V_{\rm comp}}{n_{\rm RSG}} \cdot t_{\rm RSG} \, .
\end{equation}
Therefore, on first glance, it would seem that too much interleaving has an overall detrimental effect on the performance of a quantum computer, since the additional loss due to longer delays needs to be compensated by an increased code distance, thereby increasing $V_{\rm comp}$.
Moreover, if the number of logical qubits is increased by interleaving, but these additional logical qubits remain unused in the computation, $V_{\rm comp}$ is increased by the addition of redundant volume.

However, additional qubits can typically be used to reduce the overall volume of a computation by using \textit{better-than-linear} space-time trade-offs. For instance, when measuring the volume in terms of logical qubits $\times$ $T$ gates, more advanced versions of quantum-simulation algorithms typically reduce the volume, but increase the qubit count~\cite{Reiher2017, Berry2019, vonBurg2020, Lee2020}. Such favorable trade-offs also appear on the level of subcomponents of larger algorithms, such as more efficient loading of classical data into the quantum computer by using more qubits~\cite{Low2018}. Moreover, with surface codes, magic states are required to execute non-Clifford gates such as $T$ gates, which need to be produced via magic state distillation~\cite{Bravyi2005}. More qubits can be used to produce magic states faster, facilitating the parallelization of $T$ gates. While this does not reduce the volume in terms of qubits $\times$ $T$ gates, it can reduce the total number of logical cycles and therefore the size of the fusion graph (and the volume in \textit{qubitseconds}). 

In all these cases, the linear space-time trade-off from interleaving unlocks volume reductions by enabling the use of more logical qubits, effectively \textit{increasing} the speed of the quantum computer on the level of a full algorithm, even though the speed of individual logical operations may be decreased. An additional advantage of slower logical operations is that the required speed of classical processing is reduced. Physical measurement outcomes need to be processed by a decoder before they can be interpreted as logical measurement outcomes. A decoder that is slower than the rate at which logical operations are executed will indirectly generate redundant computational volume, either by stalling the computation or by holding on to resource states for \textit{auto-corrected} rotations~\cite{Litinski2019,Gidney2019}. Increasing the interleaving ratio $L^2$ can therefore avoid this problem.

\subsection{Modified connectivity to improve logical gates}
\label{sec:modified_connectivity}

The interleaving modules introduced in the previous section allow universal quantum computing by combining an RSG with photonic delays and by using optical connections to other modules. So far, we have considered arrays and delays that result in planar logical qubits. However, minor modifications that fit within the modular framework of interleaving can enable a wide variety of other approaches to logic, including other topologies or geometries that can reduce the cost of logical computation. Delays and optical routing can be achieved with optical fiber, and so the optical links used between different interleaving modules do not need to be only between nearest neighbors in a 2D array. Similarly, delays can be of any duration, not only those that correspond to nearest-neighbor fusions in a 3D fusion graph. 

Here we discuss three modified logical operations: periodic boundary conditions, qubit routing, and unconventional geometries. We describe how they can be achieved by simple modifications to the interleaving modules of Fig.~\ref{fig:logicmodule} and by changing the connectivity between modules. We also discuss how some of these logical operations can be achieved using constructions based on the layered interleaving modules of Fig.~\ref{fig:l1modules}b.

\textbf{Periodic boundary conditions.} Square surface-code patches encode one distance-$d$ logical qubit using $d \times d$ physical qubits (or resource states). Toric codes~\cite{Kitaev2003} (Fig.~\ref{fig:periodic}a) are surface codes wrapped around a torus, i.e., with periodic boundary conditions, and encode twice as many logical qubits using the same number of physical qubits.
Periodic boundary conditions also benefit from lower logical error rates under equivalent conditions~\cite{LogicBlocksPaper}.
Periodic boundary conditions can be achieved within the interleaving framework with minor modifications. One approach is to add fibers to connect modules. Without changing the internal structure of the interleaving modules of Fig.~\ref{fig:logicmodule}, periodic boundary conditions can be achieved by simply connecting modules appropriately, as shown in Fig.~\ref{fig:periodic}b. 

Planar surface-code patches can also benefit from periodic boundary conditions. In a 1D arrangement of rectangular surface-code patches compatible with fast, arbitrary multi-qubit Pauli product measurements~\cite{Litinski2019}, as in the top half of Fig.~\ref{fig:periodic}c, each qubit effectively occupies a 2D footprint of $4d^2$. Here, qubits are encoded in rectangular patches that expose both logical Pauli operators to the ancilla patch, i.e., the width-$d$ region to the bottom that mediates multi-qubit measurements. The same 2D footprint of rectangular patches can also be used to encode two logical qubits in a six-corner patch, but the second qubit's Pauli operators will not be connected to the ancilla patch. With periodic boundary conditions as indicated by the red arrows in Fig.~\ref{fig:periodic}c, these six-corner patches can be effectively placed on a cylinder, such that both logical qubits encoded in each six-corner patch can be connected to the ancilla patch for arbitrary Pauli product measurements, thereby reducing the footprint by a factor of two.

\begin{figure*}[t]
\centering
\includegraphics[width=\linewidth]{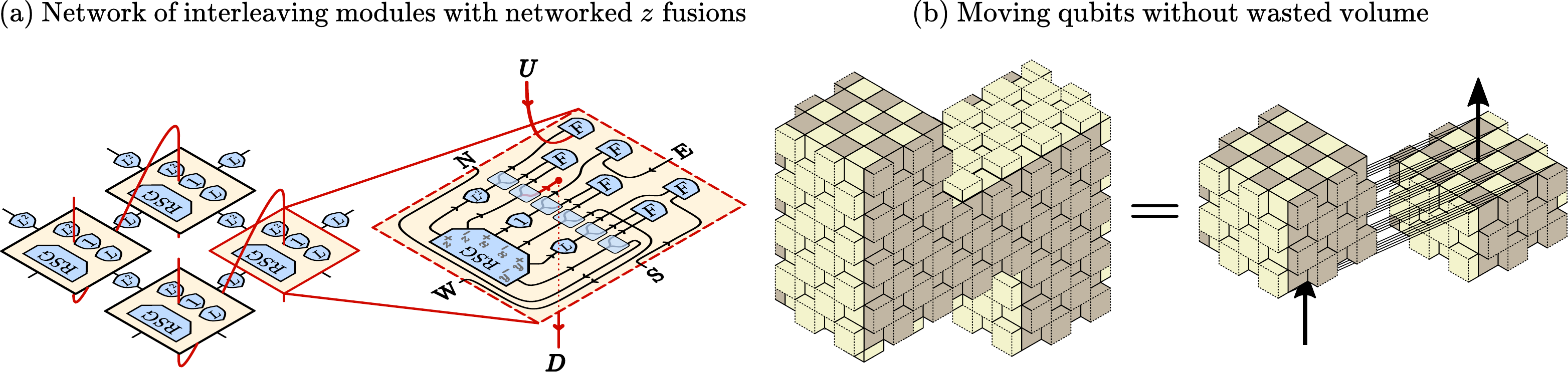}
\caption{(a) By adding a switch to the $z^\pm$-qubits, the interleaving modules of Fig.~\ref{fig:logicmodule} can be modified to include a networked $z$ fusion, i.e., $z$ fusions between resource states from different interleaving modules. These additional connections between modules are highlighted in red, where we label the additional connections $U$ (up) and $D$ (down). (b) Using networked $z$ fusions, a volume of size $\sim \! d^3$ can be removed from the logical operations that move logical qubits from one location to another.}
\label{fig:qubitmove}
\end{figure*}

An alternative approach to periodic boundaries is shown in Fig.~\ref{fig:periodic}d. Here, a single module with a 1-, $M$- and $M^2$-delay can implement a fusion graph with toric slices of size $M \times M$ without using any switches.
Note that the toric fusion graph implemented by this module includes a shift by one lattice site when wrapping around the torus.
In other words, a translation by $M$ lattice sites in the $x$ direction and one lattice site in the $y$ direction is required to map a fusion-graph vertex back to itself, or, equivalently, a translation by $M$ sites in the $y$ direction and one site in the $z$ direction. 
Similarly, toric slices of size $M_1 \times M_2$ can be implemented by replacing the $M$- and $M^2$-delay with an $M_1$- and $M_1M_2$-delay. 

For large $M$, when generating a torus much larger than a single logical qubit, the size of the longest delay can be prohibitive. A systematic way to reduce the size of the maximum delay length is shown in Fig.~\ref{fig:periodic}e. Here, a network of $n$ modules is used, where the RSGs of each module still produce one resource state every $t_{\mathrm{RSG}}$, but they are staggered, such that the RSG cycles of the $k$-th module are shifted by $(k/n)\cdot t_{\mathrm{RSG}}$. Modules are arranged on a ring with $x^\pm$-connections between nearest neighbors, $y^\pm$-connections between neighbors separated by $\lambda = M \mod n$ modules, and $z^\pm$-connections between neighbors separated by $\mu = M^2 \mod n$ modules. In such a network, the delay lengths are reduced to $1/n$, $M/n$ and $M^2/n$.

In the absence of switches, the network of modules in Fig.~\ref{fig:periodic}e is only capable of producing a toric fusion-graph structure. However, if the fusion devices of these modules are replaced with \textit{reconfigurable} fusion devices (as in Fig.~\ref{fig:logicmodule}), such modules can also execute logical operations. This presents an interesting alternative approach to logical operations, wherein logical gates are effectively carved out of a torus and all switching happens inside the reconfigurable fusion devices without the need for switches that route photons to other modules or fusion devices.

\textbf{Moving logical qubits.} 
Interleaving modules can also be reconfigured to allow more compact compiling of trivial logical operations. When restricted to a 2D array of logical qubits, there may be a significant volume cost associated with moving a logical qubit from one location to another. The fusion graph (and corresponding spacetime diagram) of such an operation is shown in Fig.~\ref{fig:qubitmove}b. This could happen, for example, when moving magic states between different stages of distillation. Since moving a logical qubit does not apply a logical operation to this qubit, the volume required for the operation can be removed or reduced by adding connections between the relevant logical qubits.

One approach to doing this with additional connections between neighboring modules is shown in Fig.~\ref{fig:qubitmove}. In these modules, we start with the interleaving modules of Fig.~\ref{fig:logicmodule}, but add the option to perform networked $z$ fusions, i.e., $z$ fusions between resource states from different interleaving modules, as shown in Fig.~\ref{fig:qubitmove}a. With this addition, a volume of size  $\sim \! d^3$ can be saved when moving a logical qubit between two neighboring sites, as shown in Fig.~\ref{fig:qubitmove}b.

More generally, connections between spatially separated interleaving modules can also help with qubit routing. In an architecture consisting of multiple different logic units, e.g., units responsible for storing data qubits and units responsible for producing non-Clifford resource states via magic state distillation, logical qubits often need to be moved from unit to unit. 
Incorporating multiple oddly-shaped units in a connected 2D architecture typically leads to wasted space, since these units cannot always be packed tightly. 
With non-local networked connections, individual units can be constructed separately and connected without any regard for packing constraints, thereby avoiding redundant operations.

\begin{figure*}[t]
\centering
\includegraphics[width=\linewidth]{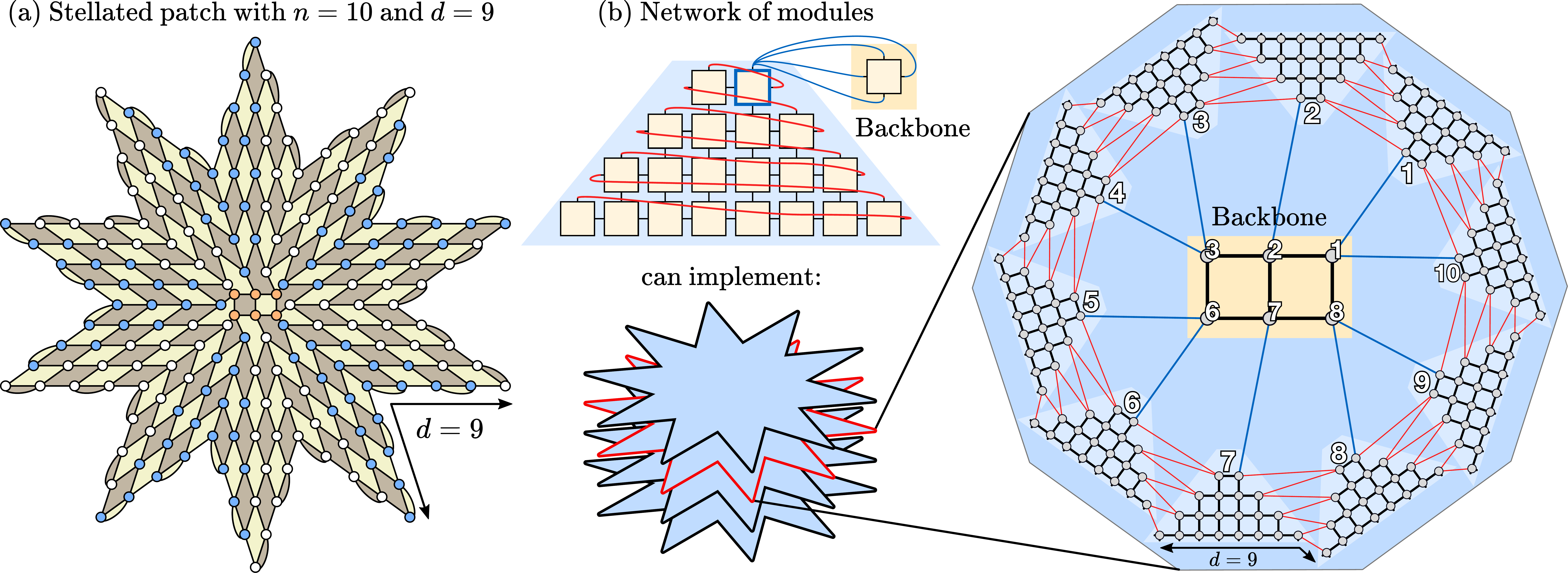}
\caption{Implementation of stellated patches. (a) Time-slice diagram of an $n=10$ stellated $n$-gon surface-code patch with code distance $d=9$. It consists of 10 triangles, of which five are highlighted in blue and five in white. In addition, there is a 6-qubit backbone highlighted in orange. (b) Network of interleaving modules with switchable network connections. The north connection of the module highlighted in blue can be switched to connect to one of the four sides of the backbone module. Furthermore, the red connections can be switched between connecting with a 1-delay or an $(n-1)$-delay. These modules produce 2D slices of a stellated $n$-gon fusion graph in $n$ RSG cycles using the indicated time-bin coordinates.}
\label{fig:stellated}
\end{figure*}

\textbf{Unconventional geometries.} 
The previous examples can all be viewed as modifications of rastered interleaving modules.
We now focus on constructions that are based on the \textit{layered} interleaving modules in Fig.~\ref{fig:l1modules}b.
Such modules can facilitate the implementation of unconventional geometries, provided these geometries involve a repeating macroscopic pattern.
We will modify these modules by adding switchable network connections.
In the networks considered so far, every interleaving module has a unique neighbor to one of the six directions.
By adding more switching options and fusion devices, one can perform networked fusions between different (but fixed) combinations of neighbors.
Such switchable network connection can also be used to select between different delay lengths and different possible fusion partners from the same neighbor.

We first consider the example of stellated surface-code patches~\cite{Kesselring2018}.
These are $n$-gon generalizations of triangular~\cite{Yoder2017} ($n=3$) and square ($n=4$) surface-code patches.
We consider a version of stellated surface-code patches with even $n$, for which a qubit-based distance-$d$ $n$-gon patch encodes $n/2-1$ logical qubits using $nd^2/4 + (3n-16)/4$ physical qubits.
Asymptotically, for large $n$, stellated surface-code patches use $d^2/2$ physical qubits per logical qubit, half compared to square patches.
However, for large $n$, the shape of stellated patches makes them difficult to implement in a regular 2D array of physical qubits.
An example of an $n=10$ patch encoding 4 qubits is shown in Fig.~\ref{fig:stellated}a.
The figure shows a time-slice diagram which can be both interpreted as a qubit-based diagram where each vertex is a physical qubit or as a 2D fusion-graph slice where each vertex is a resource state, as explained in Fig.~\ref{fig:logicoverview}. 

To implement a fusion-graph structure with layered interleaving, it is useful to identify a repeating pattern.
A stellated $n$-gon surface-code patch can be split into $n$ identical triangles containing $(d^2-1)/4$ vertices (qubits or resource states).
These triangles are connected to a \textit{backbone} consisting of $2 \times (n-4)/2$ vertices.
Slices of the corresponding fusion graph can be produced by the network of modules shown in Fig.~\ref{fig:stellated}b, which contains $(d^2-1)/4$ modules arranged in a triangle and one additional module responsible for the backbone.
The modules produce a slice of the stellated patch in $n$ RSG cycles.
If the modules arranged in a triangle were layered interleaving modules (Fig.~\ref{fig:l1modules}b) with $k=n$, they would produce $n$ disconnected triangles in each slice.
Thus, additional network connections are required to connect these triangles to a stellated patch, namely the connections highlighted in red and blue in Fig.~\ref{fig:stellated}b.
The network connections highlighted in red are used to connect triangles produced in different cycles.
Since these can either be 1 or $(n-1)$ RSG cycles apart (due to the connection between the first and the $n$-th triangle), this connection needs to be switchable between connecting via a 1-delay or an $(n-1)$-delay. 

The module highlighted in blue is connected to the backbone module as its neighbor to the north.
This connection also needs to be switchable, since the $y^-$-qubit from the blue module can either fuse with the $x^+$, $x^-$, $y^+$ or $y^-$-qubit from the backbone module, thereby adding the ability to switch between different fusion partners.
In the interleaving coordinate assignment shown in Fig.~\ref{fig:stellated}b, we assign time bins 1 through $(n-4)/2$ from right to left to the resource states in the top row of the backbone, and cycles $n/2 + 1$ through $n-2$ from left to right to those in the bottom row.
Therefore, the inter-module connections to the east and west side of the backbone module involve a 1-delay, and those to the north and south side involve no delay.
There is an additional connection to the east side of the backbone module with an $(n-1)$-delay.
Note that, in every time slice, the backbone module produces 4 redundant resource states that are not used (and therefore discarded), namely those produced in RSG cycles $n/2-1$, $n/2$, $n-1$ and $n$. 

Thus, even though  stellated $n$-gon patches with large $n$ are difficult to implement in a regular 2D lattice of physical qubits, layered interleaving
significantly facilitates their implementation, enabling a more efficient encoding of logical qubits.

\begin{figure}
    \centering
    \includegraphics[width=0.99\linewidth]{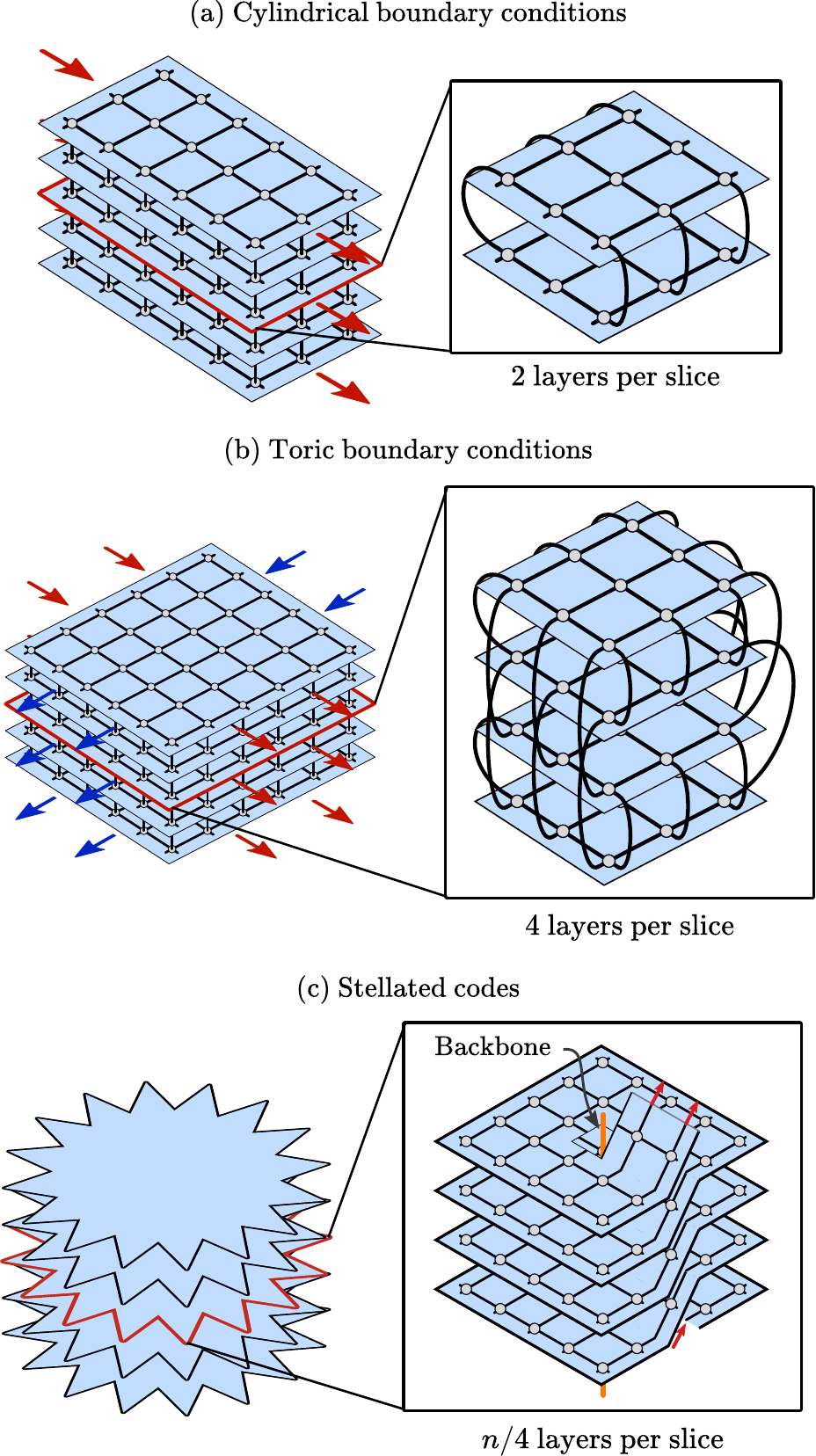}
    \caption{Folded-layer constructions of fusion graphs. (a) For cylindrical boundary conditions, each slice of the fusion graph can be folded into two layers with local inter-layer connections. (b) For toric boundary conditions, this can be done using four layers. (c) Slices of stellated $n$-gon patches with $n = 0 \mod 4$ can be folded into $n/4$ square layers.}
    \label{fig:modifying_topology_with_time}
\end{figure}

\textbf{Layered interleaving with folded-layer constructions.}
The previous example shows that fusion graphs with repeating macroscopic patterns are amenable to layered interleaving.
Other types of fusion graphs that work well with layered interleaving are folded-layer constructions. 
If a fusion graph can be decomposed into slices, and each slice can be folded into identical \textit{layers} with local inter-layer connections, modules based on the layered interleaving modules of Fig.~\ref{fig:l1modules}b can be used to implement this fusion graph layer by layer.

We show three examples of this in Fig.~\ref{fig:modifying_topology_with_time}.
In all these examples, each 2D slice of the fusion graph is folded into multiple identical layers.
In Fig.~\ref{fig:modifying_topology_with_time}a, a fusion graph with cylindrical boundary conditions is shown.
Each slice of this fusion graph can be folded into two identical layers with local inter-layer connections along two of the boundaries.
This folding construction can be used to implement periodic boundary conditions without requiring non-local connections as in Fig.~\ref{fig:periodic}.
A network of layered interleaving modules can implement this structure, if the modules include a 1-delay to create the connections at the boundaries of layers, as well as a 2-delay to connect different slices, where each slice is a pair of layers.

Similarly, slices of a fusion graph with toric boundary conditions can be folded into four layers, as shown in Fig.~\ref{fig:modifying_topology_with_time}b.
Here, each slice is created in four RSG cycles with layered modules using a 1-delay and 2-delay for inter-layer connections, as well as 4-delay for connections between slices.
Note that, in this setting, the length of the delays required to create these periodic boundary conditions is constant, and does not change if the size of the code increases. Moreover, only spatially local connections between modules are required.
This can be thought of as using interleaved layers to create a small, constant-sized, fourth dimension that allows for a local embedding of the torus.

The final example in Fig.~\ref{fig:modifying_topology_with_time}c shows a folded-layer construction of $n$-gon stellated patches.
Instead of splitting each patch into $n$ triangles, here, each patch is split into $n/4$ layers of cut squares with local inter-layer connections along the cut.
Layered interleaving modules with 1-delays and $(n-1)$-delays for inter-layer fusions, and $n$-delays for inter-slice fusions can implement this structure, but require additional modules for the implementation of the backbone.
Again, the layered construction turns the spatially non-local connections highlighted in red in Fig.~\ref{fig:stellated} into local connections.

\begin{figure}[t]
\centering
\includegraphics[width=0.95\linewidth]{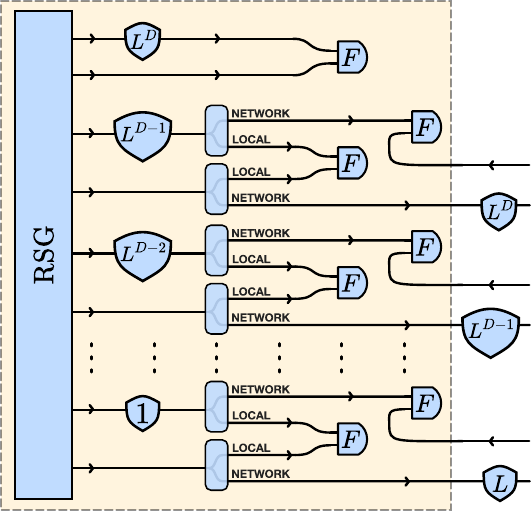}
\caption{Example of a rastered interleaving module for $D+1$-dimensional hypercubic fusion graphs using an unspecified resource state consisting of $2(D+1)$ photonic qubits. The module contains delays of lengths $L^0, L^1, \dots, L^D$ and implements the fusion graph one $D$-dimensional fusion-graph slice after the other.}
\label{fig:ddim}
\end{figure}

These examples demonstrate that layered interleaving modules are a useful tool to generate fusion graphs with repeating macroscopic patterns. However, they lack a tunable parameter similar to the rastering length that can be used to increase the interleaving ratio. Such a parameter can be introduced by combining layered with rastered interleaving. Each layer can then be produced in multiple RSG cycles using rastered interleaving, while multiple layers are combined to a fusion-graph slice using layered interleaving.

\textbf{Synthetic dimensions.} Finally, it is worth pointing out that the unconventional geometries that can be implemented by networks of interleaving modules are not restricted to 2+1-dimensional fusion graphs. Rastered interleaving modules can also be used to implement $D+1$-dimensional fusion graphs with $D>2$, such that interleaving can be used to facilitate the implementation of higher-dimensional topological codes. In this case, a different resource state will be required, as the implemented code no longer corresponds to a 2-dimensional surface code.

Suppose that the fusion graph of this $D$-dimensional code is a $D+1$-dimensional hypercubic lattice in which each vertex is a resource state consisting of $2(D+1)$ photonic qubits. This corresponds to the generalization of cubic $2+1$-dimensional fusion graphs. In this case, networks of interleaving modules implementing this fusion graph can be constructed as shown in Fig.~\ref{fig:ddim}. These interleaving modules contain delays of length $L^0, L^1, \dots, L^D$. Each module is connected to $2D$ neighboring modules. Such a network generates a $D$-dimensional slice of the $D+1$-dimensional fusion graph every $L^D$ RSG cycles.

\subsection{Hardware considerations}
\label{sec:hardware}

Our numerical results in Sec.~\ref{sec:numerics} demonstrate that the 6-ring fusion network can indeed tolerate the photon loss of a 1~km or longer fiber delay, assuming a transmission loss rate below 0.2 dB/km as provided by commercially available optical fiber for telecom wavelengths. Suppose that an RSG generates one resource state every nanosecond ($t_{\mathrm{RSG}} = $ 1 ns). With a speed of light in silica of 0.2 m/ns, it is possible to fit 5,000 photons separated by 1 ns inside a 1-km-long fiber, implying that $L^2 = 5000$. This also implies that, in this example, an interleaving module with a single RSG can generate approximately four logical qubits with a code distance of~35. In a circuit-based architecture, this would require at least $\sim \!$~5,000 physical data qubits, not including measurement ancillae.

One also needs to take into account that an interleaving ratio of $L^2 = 5000$ slows down the quantum computer by a factor of 5000. Each fusion-graph slice will now be produced in 5 $\mu$s instead of 1 ns. This is still competitive with solid-state qubit architectures, such as superconducting qubits or spin qubits, where the duration of a surface-code cycle (i.e., a time slice) is also expected to be in the microsecond range~\cite{Chen2021}. If the quantum computer needs to be faster, one can reduce the rastering length or employ one of the previously discussed space-time trade-offs.

Loss is the primary error source that serves as a useful guide, but in real hardware additional sources of error and practical considerations need to be taken into account. As well as the loss of the fiber itself, coupling from chip into fiber is an imperfect
process. Note that coupling loss does not affect all photons in the computation, but only photons that enter fiber delays, so it does not contribute to the baseline loss in Sec.~\ref{sec:numerics}.
Other notable sources of error are phase shifts that arise due to varying refractive index along the fiber, and timing mismatch due to the finite precision with which the fiber can be cut to a desired length. The former type of errors can be mitigated somewhat by encoding the quantum information in ``dual-temporal-rail'' encodings, i.e., two time bins that enter the same fiber. It is also worth noting that the temporal duration of the photonic wavepackets should not be too short (to lessen the impact of length mismatch), nor should it be long enough to severely restrict the number of photons that fit into a fiber delay.

\textbf{Hybrid matter-based/photonic architectures.} The results of the preceding sections show that the primary technological challenge of performing large-scale fault-tolerant quantum computing can be reduced to the repeated production of small (constant-sized) entangled states of photons compatible with readily-available high-performance optical delays. Networks of interleaving modules leverage the unique advantages of photons, namely the availability of a high-capacity quantum memory and the ability of photons to travel large distances. Matter-based devices may also take advantage of interleaving by acting as RSGs supplying resource states.

One advantage of photonic RSGs is that they can operate at fast clock rates, as single-photon sources may have gigahertz repetition rates. 
However, there is no requirement that the physical device generating the resource states need itself be photonic, and a hybrid system is possible where a state generated in any type of non-photonic system can be swapped into an appropriate photonic state. In principle, this is almost always possible. In practice, non-photonic systems are somewhat constrained by details of the typical interaction Hamiltonians available, in such a way that both the natural photonic frequencies and the achievable repetition rates are affected. The primary performance metrics that would affect the viability of such options are (i) the repetition rate at which the entangled state can be mapped into photons; (ii) the loss rate per output photon; and (iii) the photon's spectral purity (which affects fusion gate fidelity) and similar non-qubit photonic errors.

\begin{figure}[t]
\includegraphics[width=0.9\linewidth]{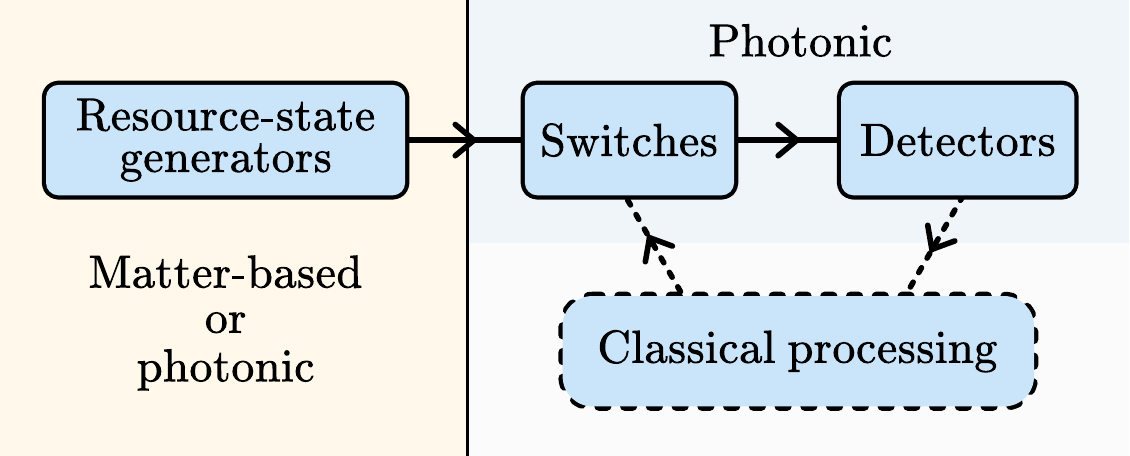}
\caption{Schematic description of a hybrid architecture. In an interleaving architecture, the main challenge is the repeated generation of identical few-qubit photonic resource states. 
This can be done by a photonic device or by a matter-based device, if the state generated within can be swapped into a photonic state. 
This provides an alternative~--~and potentially simpler~--~approach to scale up a matter-based quantum computer, since the challenge shifts from the construction of large connected arrays of physical qubits capable of universal quantum operations to the construction of small identical modules repeatedly performing the same operation without any need for complicated classical processing close to the matter-based device.}
\label{fig:hybrid}
\end{figure}

Assuming that sufficiently good performance is possible, it is worth emphasizing how much simpler a fault-tolerant quantum computer built in such a hybrid architecture (as shown in Fig.~\ref{fig:hybrid}) would be than the alternative. 
Rather than having many thousands of matter-based qubits which need to have adaptable controlled interactions, the problem is now reduced to the creation of very much smaller modules that always reproduce exactly the same state on each clock cycle. 
This photonic state can then be removed from the vicinity of the qubits (e.g., taken out of a cryostat) into a different environment, which is where all of the adaptive interactions (fusions) occur, and which is where decoding and any other strenuous classical computation can be performed. 
The homogeneous nature of the entanglement generation also obviates the need for high-throughput classical information flow to and/or from the (often cold and sensitive) location of the non-photonic qubits.

\section{Conclusion}

\textbf{Summary.} In fusion-based quantum computing (FBQC), a quantum computation is a set of fusion instructions on an arrangement of few-qubit entangled resource states.  A fusion-based quantum computer can be constructed as a network of identical modules consisting of a resource-state generator (RSG), fusion devices and a few additional linear-optical components. RSGs produce identical resource states at periodic time intervals called RSG cycles, whereas fusion devices perform entangling measurements between resource states. 
In this work, we explored how photonic memory~--~in the form of $n$-delays that store photons for $n$ RSG cycles~--~can be used to substantially increase the number of qubits available in a quantum computation. More concretely, we focused on the example of fault-tolerant FBQC with surface codes. We demonstrated that, by constructing interleaving modules containing a 1-delay, $L$-delay and $L^2$-delay, the number of qubits that a single RSG can contribute to the computational Hilbert space increases by a factor of $L^2$ due to the use of this photonic memory, albeit at the cost of decreasing the speed of logical operations by the same factor.

\textbf{Interleaving uses fiber as memory to substantially increase the number of logical qubits.} Remarkably, $L^2$-delays with $L^2>1000$ can be constructed with available technology. For RSG cycles on the order of nanoseconds, kilometer-long low-loss optical fiber with telecom-wavelength transmission loss below 0.2 dB/km can act as a photonic memory for thousands of photons. 
RSGs equipped with such a high-capacity memory effectively provide the computational power of thousands of physical qubits for the purpose of fault-tolerant surface-code quantum computation. 
For nanosecond RSG cycles, interleaving ratios of $L^2 \sim 1000$ yield microsecond surface-code cycles, which is competitive with solid-state architectures.

\textbf{Interleaving reduces the cost of logical operations.} The hardware-based linear space-time trade-off from interleaving plays well with software-based trade-offs that can be better than linear. Additional qubits can be used to parallelize logical operations, e.g., by parallelizing magic state distillation, or to execute variants of algorithms for a given problem that use more qubits to decrease the overall volume of the computation in terms of qubits $\times$ operations. In this case, interleaving may decrease the speed of logical operations, but \textit{increase} the speed of the overall quantum computation. If such trade-offs are not available, the speed of the quantum computer can still be increased by connecting more modules to the network. Moreover, since different interleaving modules are connected via macroscopic optical fiber, these modules need not be arranged in a two-dimensional grid with strict nearest-neighbor connectivity. We have shown a few basic examples demonstrating that modified and switchable connections can reduce the cost of various logical operations, and hope that this will be used as a toolbox to find many additional improvements in the future.

\textbf{Networks of interleaving modules leverage the unique advantages of photons}, namely the availability of a high-capacity quantum memory and the ability of photons to travel large distances. 
While RSGs can be constructed using linear-optical components~\cite{GimenoSegoviaThesis} in a purely photonic architecture, the advantages of interleaving can also be used by non-photonic matter-based systems, if they can be made to function as resource-state generators. 
This suggest that interleaving provides an alternative approach to scale up matter-based qubits. 
Instead of constructing connected million-qubit arrays of physical qubits, interleaving can turn arrays of disconnected few-qubit devices into a large-scale quantum computer, provided that their qubits can be converted to photonic qubits. 
In such a hybrid architecture, the matter qubits are solely responsible for the repeated generation of identical few-qubit resource states, whereas the photonic components connect them to a large-scale fault-tolerant quantum computer, provide high-capacity memory for interleaving, and handle the classical processing associated with error correction. 
With such an approach, the problem of constructing a large-scale fault-tolerant quantum computer is reduced to the construction of many identical and autonomously operating few-qubit resource-state generators.

\section*{Acknowledgments}

The authors would like to thank Chris Dawson and Jordan Sullivan for providing support with numerical simulations, Eric Dudley, Gabriel Mendoza, Chris Sparrow and Mihai Vidrighin for discussions about the limitations of hardware components, Nikolas Breuckmann for useful discussions about layered topological memory, Daniel Dries for his thorough review and helpful comments on the manuscript, and
Sara Bartolucci,
Patrick Birchall,
Hugo Cable,
Axel Dahlberg,
Andrew Doherty,
Megan Durney,
Mercedes Gimeno-Segovia,
Eric Johnston,
Konrad Kieling,
Ye-Hua Liu,
Ryan Mishmash,
Sam Morley-Short,
Andrea Olivo,
Sam Pallister,
William Pol, 
Karthik Seetharam, 
Jake Smith,
Andrzej P\'erez Veitia,
and all our colleagues at PsiQuantum for useful discussions.

\appendix

\section{Simulation methods}
\label{app:simulation}

\subsection{6-ring fusion network}
\begin{figure}[t]
    \centering
    \includegraphics[width=\columnwidth]{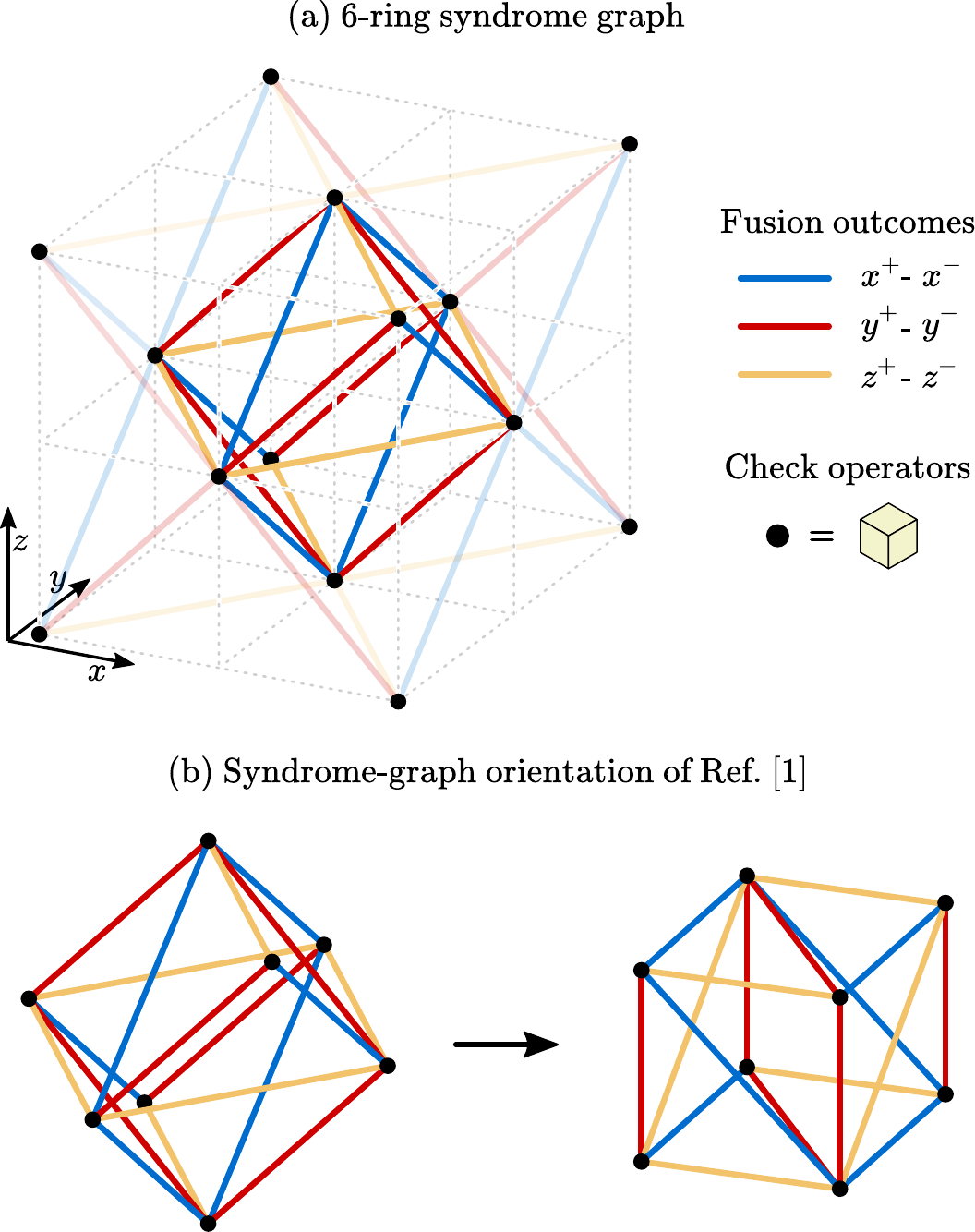}
    \caption{(a) Syndrome-graph orientation used in this paper, 18 edges of which are highlighted. (b) Transformation to the the cell-complex orientation described in Ref.~\cite{FBQCpaper}.}
    \label{fig:syndrome_compare}
\end{figure}

The example photonic fusion network that we study in Sec.~\ref{sec:numerics} is the the 6-ring fusion network introduced in Ref.~\cite{FBQCpaper}, and depicted in Fig.~\ref{fig:fusiongraph}. We note that in Ref.~\cite{FBQCpaper} the fusion network is presented in a different orientation relative to our presentation here where resource states are arranged on a cubic lattice. Figure~\ref{fig:syndrome_compare} shows the transformation used to map between the two variations. Figure~\ref{fig:syndrome_compare}a shows a repeating cell of the syndrome graph that we evaluate here. Each black circle represents a check operator and each colored edge represents a fusion measurement outcome. The dotted lines indicate the underlying cubic structure, where there is one resource state for each cubic cell. A subset of the syndrome graph edges are highlighted to show a unit cell of the syndrome graph as it is more commonly represented when analyzing a 2+1D surface code. 
When performing simulations of the fusion network, we create a large syndrome graph structure by tiling the unit cells of Fig.~\ref{fig:syndrome_compare}a with periodic boundary conditions. We define the size of such an array as $L\times L \times L$ if each side has $L$ cubes. This structure has 3 logical membranes in both the primal and dual lattice, one in each of the 3 periodic dimensions. A logical error occurs when there is an error chain that crosses the syndrome graph and crosses a periodic boundary, i.e. a non-trivial cycle on the syndrome graph.

\subsection{Linear optical fusion}
\label{app:fusion_error_model}

The fusion operation in linear optics can be viewed as a probabilistic Bell measurement which attempts to measure input qubits $q_1$ and $q_2$ in the basis $X_1X_2, Z_1Z_2$. There are three types of outcomes of a fusion on two qubits: 

\begin{enumerate}
    \item {\bf Success.} Fusion ``succeeds" with probability $1-p_{\rm{fail}}$, measuring the input qubits in the Bell stabilizer basis $ X_1X_2, Z_1Z_2$ as intended.
    \item {\bf Failure.} The fusion ``fails" with probability $p_{\rm{fail}}$, in which case it performs separable single qubit measurements. The fusion measures $Z_1I_2, I_1Z_2$ (or $X_1I_2, I_1X_2$ depending on a choice in the fusion circuit) when it fails. Therefore, one of the two desired outcomes ($Z_1Z_2$ or $X_1X_2$) can be obtained by multiplying two single-qubit measurement outcomes. Here, we randomize fusions so that on failure there is an equal probability of measuring $X_1X_2$ or $Z_1Z_2$.
    \item {\bf Loss.} If a photon is lost, a fusion detects the wrong number of photons. In this case neither intended stabilizer outcome is measured.
\end{enumerate}

We use a boosted type-II fusion~\cite{grice2011arbitrarily}, which uses an ancillary Bell pair and has $p_{\rm{fail}} = 1/4$. This fusion operation acts on four photons in total, two input photons and two photons in the Bell pair.
One of the input photons and both boosting photons are fused immediately after they are created, and so experience a loss of $p_{\rm baseline}$. One of the photons goes through an additional fiber delay which suffers a total loss of $p_{\rm interleaved}(N) = 1- (1-p_{\mathrm{baseline}})(1-p_{\text{\clock}})^{N} $, where $N$ indicates the number of time bins in the delay, and $p_{\text{\clock}}$ is the loss per time bin due to propagation in interleaving fiber.
Fusion succeeds only when no photon is lost, which happens with probability $\eta_f = (1-p_{\rm interleaved})(1-p_{\rm baseline})^3$. The erasure probability for a fusion measurement in the absence of loss is $p_{\rm{fail}}/2 = 1/8$. Consequently, the marginal erasure probability for each individual measurement is

\begin{equation}
p_0 (N, p_{\rm baseline}) = 1 - \frac{7}{8}(1-p_{\text{\clock}})^N(1-p_{\rm baseline})^4 \, ,
\label{eq:p0_fusion_meas_erase_prob}
\end{equation}

\noindent which we call the physical fusion measurement erasure probability.

\subsection{Encoded fusion erasure probability}
\label{subapp:encoded_fusion_erasure_prob}

We use encoded fusion in the fusion network, where each qubit in the resource state is encoded in a four-qubit Shor code which is a concatenation of two two-qubit repetition codes. There are two variants of this code, which we refer to as $C_1$ and $C_2$. $C_1$ has stabilizers $\langle X_1X_2X_3X_4, Z_1Z_3,Z_2Z_4 \rangle$ (X repetition above Z) while $C_2$ has stabilizers $\langle Z_1Z_2Z_3Z_4,X_1X_3,X_2X_4\rangle$ (Z repetition above X).

Given two encoded qubits $A$ and $B$, we perform an encoded fusion by performing pairwise fusions transversally, giving two ways to reconstruct each of the encoded fusion outcomes. 
For code $C_1$, the erasure probability of  $\overline{X_A}\overline{X_B}$ is $[1 - (1-p_0)^2]^2$, and the erasure probability of  $\overline{Z_A}\overline{Z_B}$ is $1 - (1-p_0^2)^2$. For code $C_2$, these probabilities are swapped between the two encoded fusion outcomes. We choose between the encodings $C_1$ and $C_2$ randomly at each site, such that the average encoded fusion erasure probabilities are uniformly given by

\begin{equation}
p_{\rm{enc}}(p_0) = \frac{[1 - (1-p_0)^2]^2 + 1 - (1-p_0^2)^2}{2} \, .
\label{eq:penc_fusion_meas_erase_prob}
\end{equation}
Note that $p_{\rm{enc}} < p_0$ when $p_0 < 0.5$.

\begin{figure}[t]
 \centering
    \includegraphics[width=\linewidth]{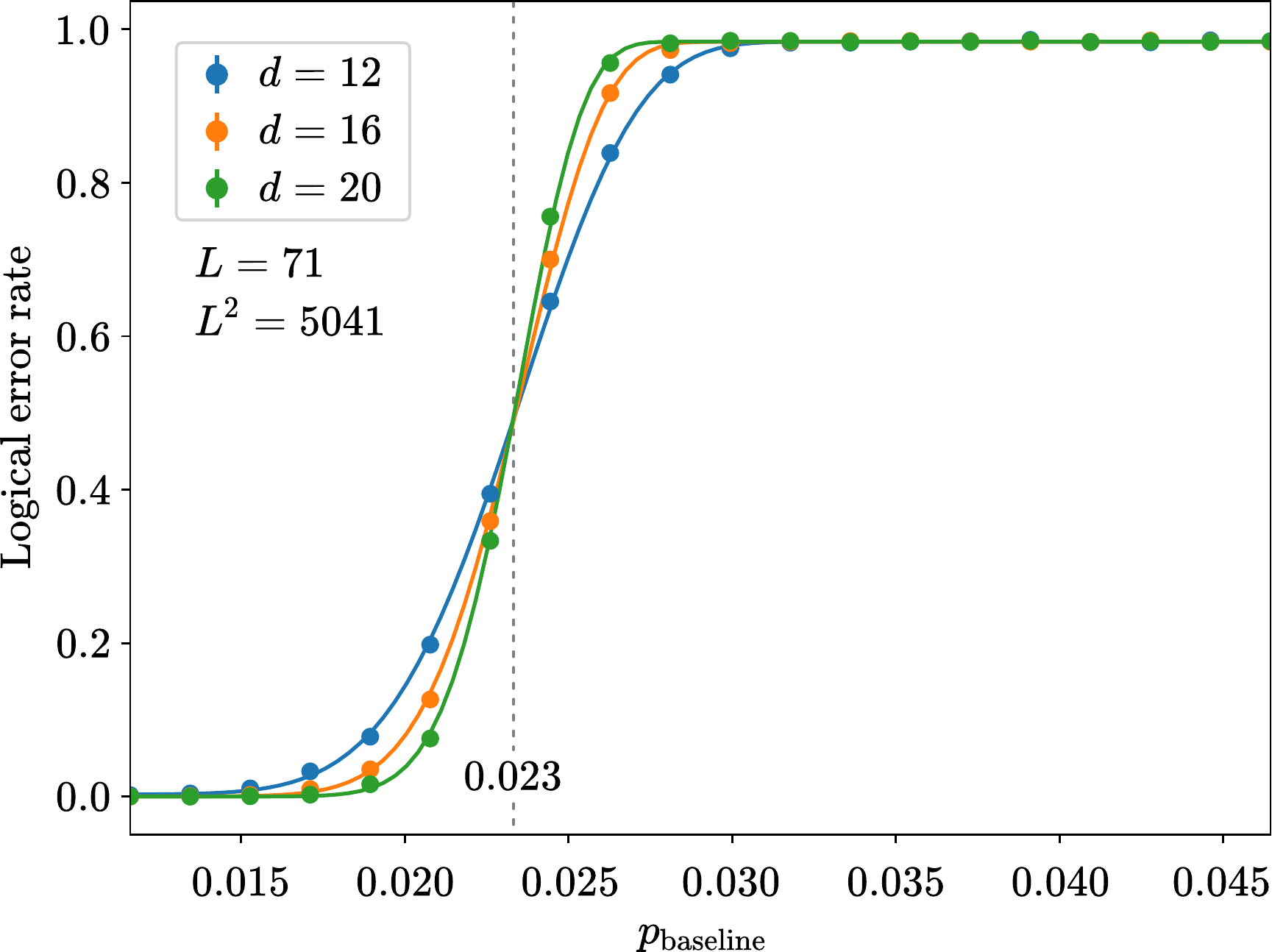}
    \caption{Data identifying the threshold against baseline loss $p_{\rm baseline}$ in the case that the maximum delay length is $L^2=5041$ time bins.}
    \label{fig:interleavingthresholdplot_Li71}
\end{figure}

\subsection{Numerical methods}
\label{sec:simulation_methods}

Here, we describe the methods used to obtain the numerical results shown in Fig.~\ref{fig:loss_threshold}.

\textbf{Error model.} We consider the 6-ring fusion network with a (2,2)-Shor encoding, where each fusion in the network performs the encoded fusion described in Sec.~\ref{subapp:encoded_fusion_erasure_prob}. Each photon in the resource state experiences an independent loss of $p_{\rm loss} = 1- (1-p_{\rm baseline})(1-p_{\text{\clock}})^N$ prior to its fusion, where $p_{\rm baseline}$ is the loss due to state preparation, $N$ indicates the number of time bins in the delay, and $p_{\text{\clock}}$ is the loss per time bin due to propagation in the interleaving fiber. $N_{1-6} = \{0, 0, 0, 1, L, L^2\}$ for each the six resource state qubits. Each of the four photons in each encoded qubit undergoes the same delay. The encoded fusion erasure probability for each measurement is calculated using Eq.~\ref{eq:p0_fusion_meas_erase_prob} and 
\ref{eq:penc_fusion_meas_erase_prob} with $p_{\rm interleaved} = 1  - (1-p_{\rm baseline})(1-p_{\text{\clock}})^{L^i}$, $i = 0, 1, 2$ for encoded fusions in the $x$, $y$ and $z$ directions respectively. The encoded fusion erasure probability for each measurement is therefore given by: 

\begin{align}
    p_{\rm x,erasure} &= p_{\rm enc}(p_0(1, p_{\rm baseline})) \\
    p_{\rm y,erasure} &= p_{\rm enc}(p_0(L, p_{\rm baseline}))\\
    p_{\rm z,erasure} &=p_{\rm enc}(p_0(L^2, p_{\rm baseline}))
\end{align}

\textbf{Computing the logical error rate.} For each combination of error parameters we simulate a 3D block of the fusion network with periodic boundary conditions of size $d~=~\{12,16,20\}$ resource states in each dimension (i.e., the number of vertices along one side of the fusion graph). To perform a single numerical trial, we sample erasures on the edges of the primal and dual syndrome graph with the probabilities stated above.

We then perform decoding using the union-find decoder~\cite{unionfind} on the syndrome graph. Since we are using a loss-only model, this decoder is optimal and is reduced to the limiting case of simply performing the peeling decoder of Ref.~\cite{Delfosse2020}. Running the decoder results in a correction, which we compare against the error sample to identify whether a logical erasure is introduced. We decode primal and dual syndrome graphs separately, and identify the logical error state of both after decoding. We say that a logical error is introduced, if either the primal or dual syndrome graph has a logical error in any one of the three periodic dimensions. We repeat this sampling and decoding process at least 15,000 times for each combination of error parameters to compute the logical error rate. 

\textbf{Finding the threshold.} To identify a threshold we fix the delay length parameter $L$. We then sweep the value of $p_{\rm baseline}$, computing the logical error rate for each value, and fit to the cumulative distribution function (CDF) of the rescaled and shifted beta distribution to identify a threshold crossing. Figure \ref{fig:interleavingthresholdplot_Li71} shows such a sweep (points) and fit (lines) for the 6-ring fusion network with $L=71$. The crossing of the different curves here allows us to estimate the threshold for the fixed value of $L$.

\textbf{Baseline threshold as a function of delay length}. Each threshold calculation identifies the baseline threshold for a single value of $L$. Repeating this for a series of values of $L$, we produce the data shown in Fig.~\ref{fig:loss_threshold}b.

\end{document}